\documentclass[]{jfm}

\usepackage{graphicx}
\usepackage{newtxtext}
\usepackage{newtxmath}
\usepackage{natbib}
\usepackage{hyperref}
\hypersetup{
    colorlinks = true,
    urlcolor   = blue,
    citecolor  = black,
}

\usepackage{multirow}
\usepackage[labelformat=simple,font=it]{subcaption}

\usepackage[normalem]{ulem}

\newcommand\We{\mbox{\textit{We}}}
\newcommand\Oh{\mbox{\textit{Oh}}}

\newcommand{\RomanNumeralCaps}[1]
\linenumbers

\newcounter{aqctr}
\newenvironment{author-query}
{\refstepcounter{aqctr}\par\vspace{\baselineskip}\noindent
\color{red}\textbf{Author Query/Comment AQ \arabic{aqctr}.}}
{\par\vspace{\baselineskip}\normalcolor}

\title{
Correlation between morphological evolution of splashing drop and exerted impact force revealed by interpretation of explainable artificial intelligence
}

\author{Jingzu Yee\aff{1},
  Daichi Igarashi\aff{1},
  Pradipto\aff{1},
  Akinori Yamanaka\aff{1}
 \and Yoshiyuki Tagawa\aff{1,2}
   \corresp{\email{tagawayo@cc.tuat.ac.jp}}}

\affiliation{\aff{1}Department of Mechanical Systems Engineering, Tokyo University of Agriculture and Technology, 2-24-16, Naka-cho, Koganei, Tokyo 184-8588, Japan
\aff{2}Institute of Global Innovation Research, Tokyo University of Agriculture and Technology, 2-24-16, Naka-cho, Koganei, Tokyo 184-8588, Japan}

\begin{document}
\maketitle

\begin{abstract}
This study reveals a possible correlation between splashing morphology and the normalized impact force exerted by an impacting drop on a solid surface. 
This finding is obtained from a newly proposed feature extraction method and a subsequent interpretation of the classification of splashing and non-splashing drops performed by an explainable artificial intelligence (XAI) video classifier.
Notably, the values of the weight matrix elements of the XAI that correspond to the extracted features are found to change with the temporal evolution of the drop morphology.
We compute the rate of change of the contributions of each frame with respect to the classification value of a video as an important index to quantify the contributions of the extracted splashing and non-splashing features at different impact times to the classification of the XAI model.
Remarkably, the rate computed for the extracted splashing features is found to closely match the profile of the normalized impact force, where the splashing features are most pronounced immediately after the normalized impact force reaches its peak value.
This study has provided an example that clarifies the relationship between the complex morphological evolution of a splashing drop and physical parameters by interpreting the classification of an XAI video classifier.
\end{abstract}

\begin{keywords}
Authors should not enter keywords on the manuscript, as the author must choose these during the online submission process and will then be added during the typesetting process (see \href{https://www.cambridge.org/core/journals/journal-of-fluid-mechanics/information/list-of-keywords}{Keyword PDF} for the full list).  Other classifications will be added at the same time.
\end{keywords}

{\bf MSC Codes }  {\it(Optional)} Please enter your MSC Codes here

\section{\label{sec:intro}Introduction}

The impact of a liquid drop on a solid surface is a high-speed phenomenon that is encountered in a variety of contexts, such as spray cooling \citep{breitenbach2018drop} and aircraft icing \citep{lavoie2022penalization, zhang2016effect}.
Under certain impact conditions, splashing can occur, i.e. the impacting drop breaks up and ejects secondary droplets~\citep{gordillo2019note, hatakenaka2019magic, yokoyama2022droplet} instead of just spreading over the surface until it reaches its maximum radius~\citep{gordillo2019theory, clanet2004maximal}.
Splashing has various consequences, such as soil erosion \citep{fernandez2017splash},  propagation of contaminants \citep{gilet2015fluid, waite2015grapevine}, and visible decreases in printing and painting quality \citep{lohse2022fundamental}.
Therefore, it is necessary to understand the dynamics of a splashing drop from the morphological evolution that occurs during the impact process.
Owing to the multiphase nature of this phenomenon, which involves the liquid drop, the solid surface, and the ambient air, many physical parameters strongly influence the occurrence of splashing \citep{josserand2016drop, rioboo2001outcomes, yarin2006drop}.
For instance, a given parameter can either promote or suppress splashing, depending on other parameters \citep{usawa2021large, zhang2022surface, zhang2021reversed}.
Furthermore, the spreading dynamics of a splashing drop are very complex, because the ejected secondary droplets add more morphological features, such as their ejection angle~\citep{burzynski2020splashing}, ejection velocity~\citep{thoroddsen2012micro, mundo1995droplet}, number~\citep{lin20223d}, and size \citep{juarez2012splash, riboux2015diameters, wang2018unsteady}.

To aid observations of splashing drops,  attention has turned to artificial intelligence (AI), which has been widely adopted and has proved effective in carrying out tasks in different fields, such as image and video processing~\citep{he2016deep, krizhevsky2012imagenet, voulodimos2018deep}, aeronautical and aerospace engineering~\citep{brunton2021data, hou2019machine, li2020efficient}, and fluid mechanics~\citep{brunton2020machine, colvert2018classifying, erichson2020shallow}.
Although the underlying reasoning that leads AI to a specific decision is often unknown or not correctly understood~\citep{adadi2018peeking, arrieta2020explainable}, by solving the problems of explainability and interpretability, AI can become a powerful tool for advancing knowledge of physical phenomena.
In particular, in studies of turbulence, AI has been widely used for the reconstruction of turbulence fields \citep{fukami2019super, kim2021unsupervised}, for inflow turbulence generation in numerical simulations \citep{kim2020deep, yousif2023transformer}, and to gain physical insight from data \citep{kim2023interpretable, kim2020prediction, lu2020extracting}.
On the other hand, the application of AI to the investigation of multiphase flows is relatively recent.
With regard to drop impact,  several AI-based studies have been published since 2021 on the prediction of post-impact drop morphology \citep{yee2023prediction},  impact force \citep{dickerson2022predictive},  maximum spreading \citep{tembely2022machine, yancheshme2022dynamic, yoon2022maximum}, and splashing threshold \citep{pierzyna2021data}.
Notably, by image feature extraction using explainable artificial intelligence (XAI), \citet{yee2022image} observed that the contour of a splashing drop's main body is higher than that of a non-splashing drop.
Although \citet{yee2022image} established a foundation for feature extraction methodology using XAI, the relationship between morphological features and physical parameters has not been discovered.
This is because the classification was conducted solely on a single snapshot at a specific impact time, which does not enable comparison of morphological evolution between splashing and non-splashing drops.
However, because the temporal evolution of the morphology is related to the acceleration of the drop, it contains important information about physical parameters, such as the impact force.

In the present study, a classification of videos or image sequences is proposed, on the basis of which the morphological evolution of splashing drops can be compared with that of non-splashing drops.
Although recurrent neural networks (RNNs) and long short-term memory (LSTM) networks are the two types of AI that are most widely used to process sequential data such as audio and video data~\citep{, guera2018deepfake, ma2019ts, sherstinsky2020fundamentals}, their complex architectures cause difficulties when attempts are made to analyse their decision-making processes.
Instead, a feedforward neural network (FNN) model has been developed as an XAI video classifier, comprising a single fully connected layer.
Classification is performed on image sequences processed from high-speed videos of splashing and non-splashing drops recorded during an experiment.
The methodology of this study,  including descriptions of the dataset and the implementation of the FNN, is explained in \S~\ref{sec:method}.
After high accuracy has been attained in the classification of image sequences of splashing and non-splashing drops, an analysis of the FNN's classification process is performed to extract the features of the splashing and non-splashing drops.
An importance index is introduced to quantify the contributions of the extracted features to the classification of the FNN model.
These results and a discussion of the relationship between morphological features and physical parameters of a drop impact are presented in \S~\ref{sec:result}.
In addition, the critical impact time when the morphological differences are most pronounced is identified.
The conclusions of this study are presented in \S~\ref{sec:conclusion}

\section{\label{sec:method}Methodology}

In this section, the dataset of image sequences showing the temporal evolution of drop morphology during impact (\S~\ref{sec:dataset}) and the implementation of the FNN developed for image-sequence classification (\S~\ref{sec:ann}) are explained.

\subsection{\label{sec:dataset} Experiment setup and image processing}
\begin{figure}
\centering
\includegraphics[width=0.5\textwidth]{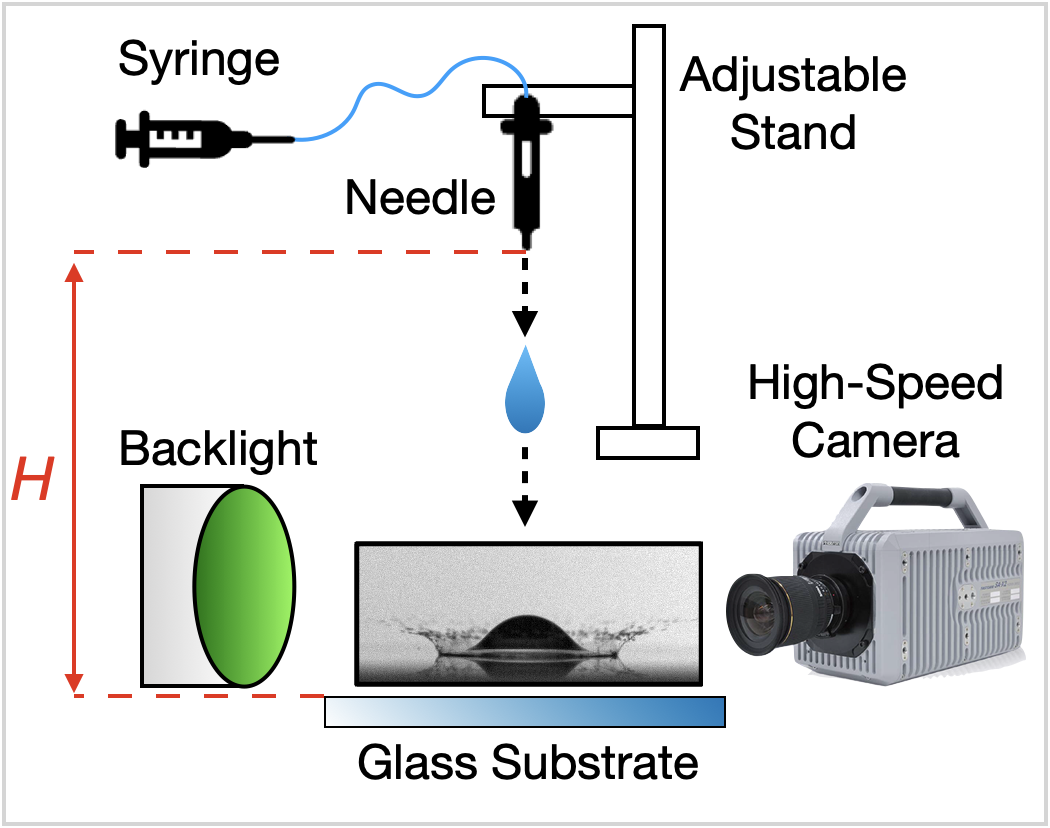}
\caption{\label{fig:setup} Schematic of  experimental setup used to collect high-speed videos of drop impact.}
\end{figure}

With the experimental setup shown in figure~\ref{fig:setup}, videos of drop impact were collected using a high-speed camera (Photron, FASTCAM SA-X) at a rate of $45\,000\,\mathrm{s}^{-1}$ and a spatial resolution of $(1.46 \pm 0.02) \times 10^{-5}\,\mathrm{m\,px}^{-1}$.
Each of the videos shows an ethanol drop (Hayashi Pure Chemical Ind., Ltd; density $\rho = 789\,\mathrm{kg\,m^{-3}}$, surface tension $\gamma = 2.2 \times 10^{-2}\,\mathrm{N\,m^{-1}}$, and dynamic viscosity $\mu = 1.0 \times 10^{-3}$\,Pa\,s) impacting on the surface of a hydrophilic glass substrate (Muto Pure Chemicals Co., Ltd, star frost slide glass 511611) after free-falling from a height $H$ ranging from $0.04$\,m to $0.60$\,m.
The area-equivalent diameter of the drop, which was measured before impact, was $D_{0} = (2.59 \pm 0.10) \times 10^{-3}$\,m.
The impact velocity $U_0$ and Weber number $\We$ ($= \rho U_{0}^2 D_{0}/\gamma$) ranged between $0.82\,\mathrm{m\,s^{-1}}$ and $3.18\,\mathrm{m\,s^{-1}}$ and between $63$ and $947$, respectively.
The splashing thresholds in terms of impact height and Weber number were $H = 0.20$\,m and $\We = 348$, respectively.
Note that some impacting drops with $H$ or $\We$ equal to or greater than the splashing threshold did not splash.
The $H$ and $\We$ of the non-splashing drop with the highest values of $H$ and $\We$ were $H = 0.22$\,m and $\We = 386$, respectively.
Thus, there was a splashing transition at $0.20\,\mathrm{m} \leq H \leq 0.22$\,m or $348 \leq \We \leq 386$.
After a frame-by-frame inspection for the presence of secondary droplets by human eyes, each of the videos was labelled according to the outcome: splashing or non-splashing.
There are a total of 249 videos: 141 of splashing drops and 108 of non-splashing drops.
\begin{figure}
\centering
\subfloat[]{
\includegraphics[width=0.45\textwidth]{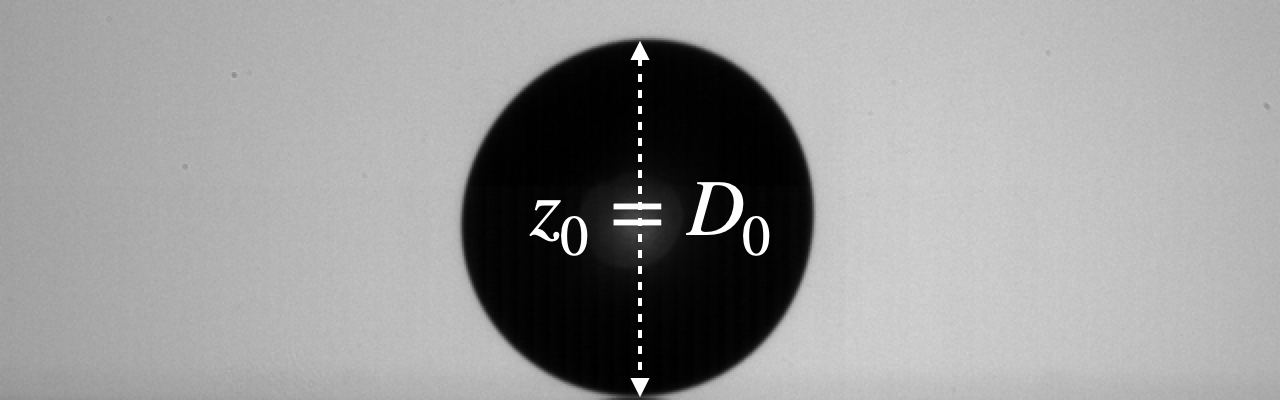}
\label{fig:z0_2R0_100}
}
\subfloat[]{
\includegraphics[width=0.45\textwidth]{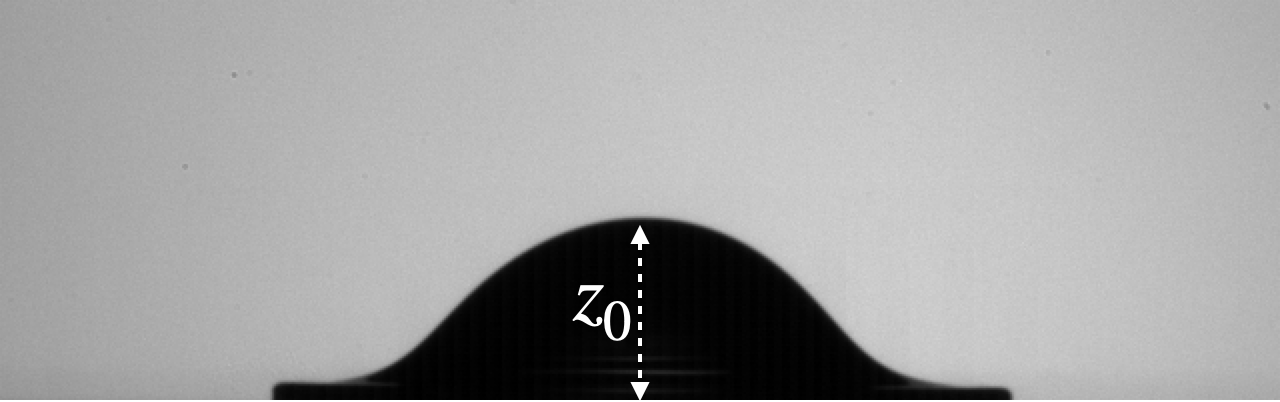}
\label{fig:z0_2R0_050}
}
\caption{\label{fig:z0_2R0}
Apex of an impacting drop: (\textit{a}) at the start of the impact when $z_0/D_0 = 1.000$; (\textit{b}) during the impact when $z_0/D_0 = 0.500$.
}
\end{figure}

\begin{figure}
\centering
\includegraphics[width=\textwidth]{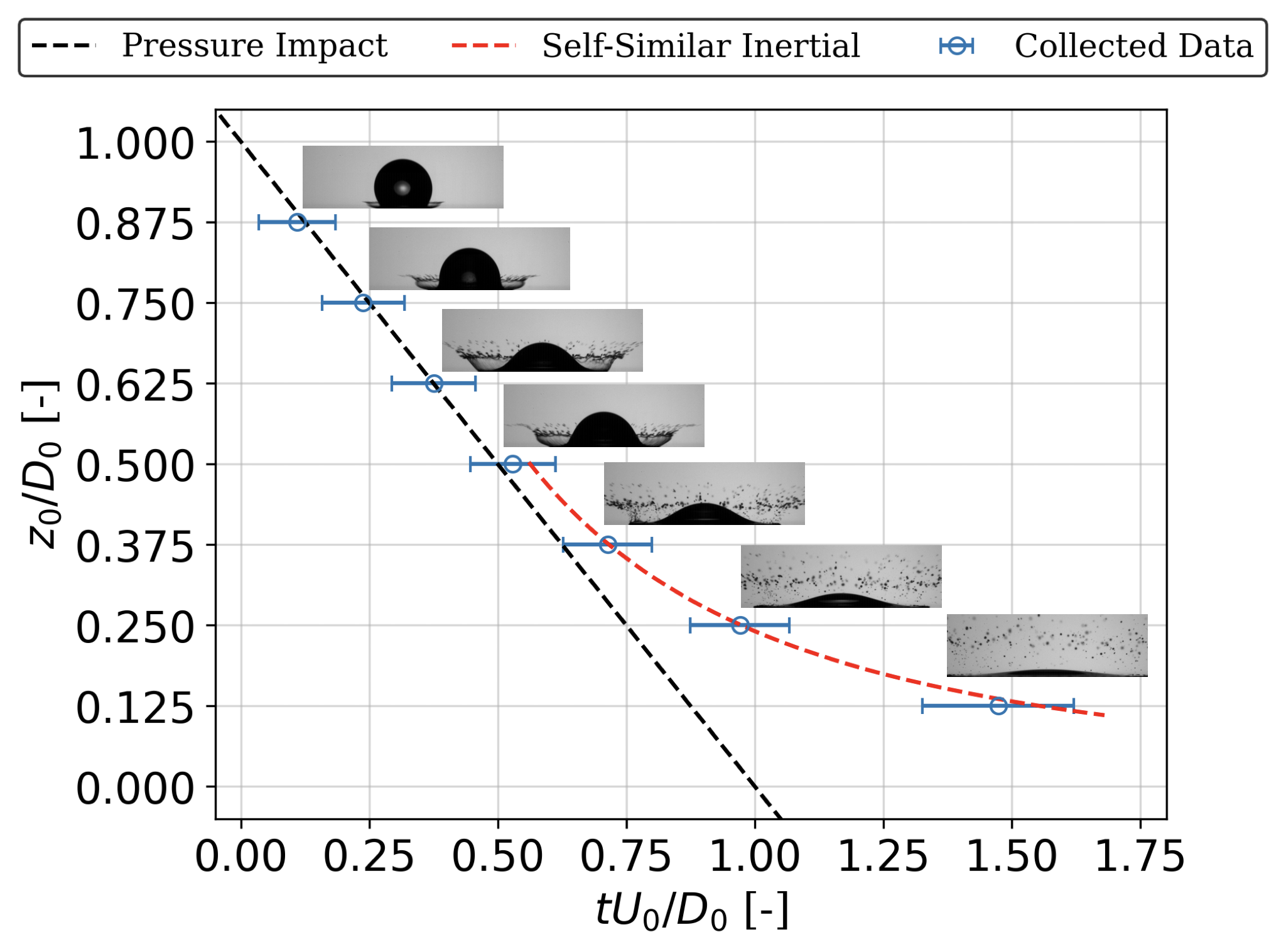}
\caption{\label{fig:apex_time_inset} 
Normalized drop apex versus normalized impact time averaged among all collected data.
}
\end{figure}

From each video, seven frames, showing the temporal evolution of the drop morphology from the start of the impact until before the drop collapsed into a pancake-like morphology, were extracted to form the image sequences for classification.
The seven frames were extracted when the normalized drop apex $z_0/D_0 = 0.875$, 0.750, 0.625, 0.500, 0.375, 0.250, and 0.125, respectively.
The definition of the drop apex $z_0$ is illustrated in figure~\ref{fig:z0_2R0}.
Therefore, $z_0/D_0$ can be understood as the portion of the drop that has yet to impact the surface.
For example, when $z_0/D_0 = 0.250$, the remaining one-quarter of the drop has yet to impact the surface.
 $z_0/D_0$ is plotted against the normalized impact time $tU_0/D_0$, which was averaged among all collected data in figure~\ref{fig:apex_time_inset}.
The error bars show the standard deviation of all 249 data.
The black and red dashed lines show the pressure impact and self-similar inertial regimes proposed by \citet{lagubeau2012spreading}, which are plotted using the equations
\begin{align}
z_0/D_0& = 1 - tU_0/D_0
\label{eq:PI},
\\
z_0/D_0 &= A_1 / (tU_0/D_0 + A_2)^2
\label{eq:SSI},
\end{align}
respectively, where $A_1$ and $A_2$ are fitting parameters, which are 0.492 and 0.429, respectively, according to the fitting results by \citet{lagubeau2012spreading}.
As shown in the figure, $z_0/D_0 = 0.875$, 0.750, and 0.625 cover the pressure impact regime; $z_0/D_0 = 0.500$ lies on the transition between the pressure impact and self-similar inertial regimes; and $z_0/D_0 = 0.375$, 0.250, and 0.125 cover the self-similar inertial regimes.

\begin{figure}
\centering
\includegraphics[width=\textwidth]{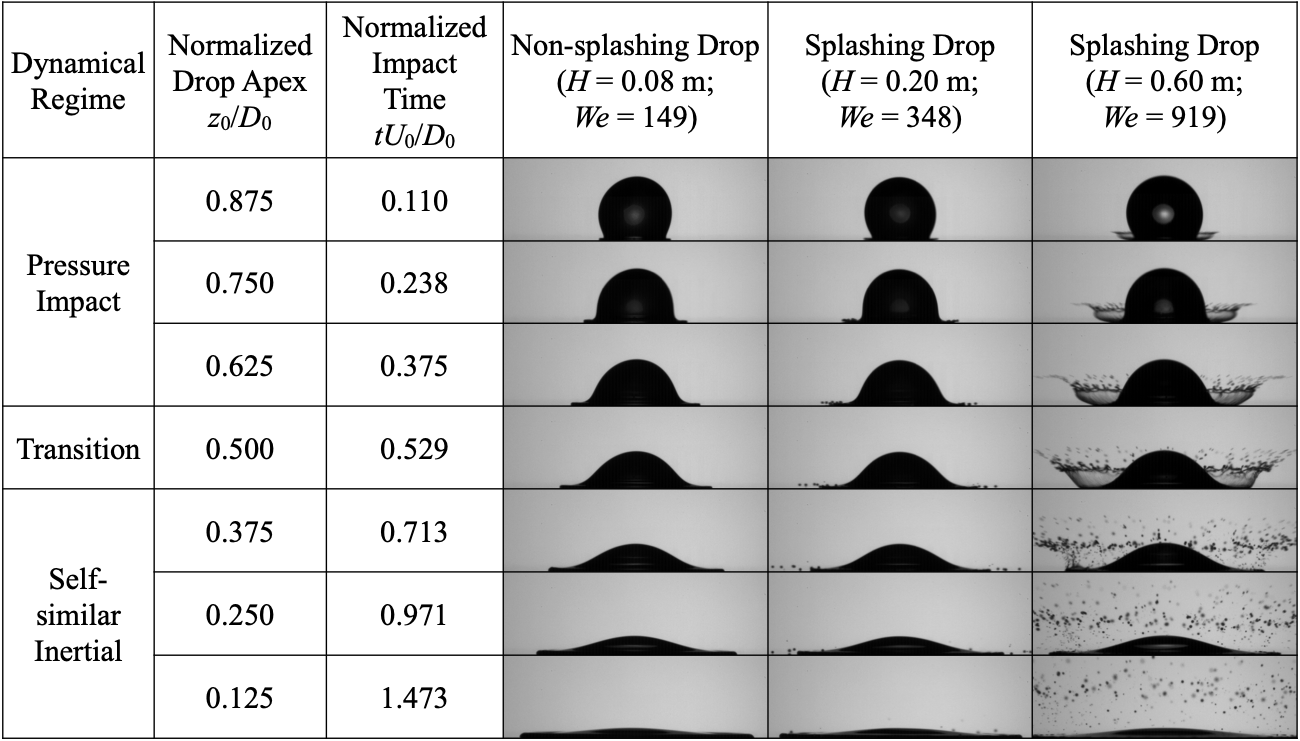}
\caption{\label{fig:processed_img} 
Examples of image sequences of splashing and non-splashing drops combined from seven frames at different normalized impact times.
}
\end{figure}

An in-house MATLAB code was utilized to extract these seven frames and to trim the background so that the impacting drop was located at the centre of each frame.
Several examples of the image sequences, including a non-splashing drop with $H = 0.08$\,m and $\We = 149$, a splashing drop at the splashing threshold with $H = 0.20$\,m and $\We = 348$, and a splashing drop with $H = 0.60$\,m and $\We = 919$, are shown in figure~\ref{fig:processed_img}.

Fivefold cross-validation was performed to ensure the generalizability of the trained FNN.
For this, the image sequences of each $H$ were segmented into five combinations of training--validation and testing data in a ratio of 80:20 
 to ensure that the data of each $H$ were included in both training--validation and testing and were distributed evenly among the data combinations.
This is reflected in the similar numbers of splashing and non-splashing data for training--validation or testing among all data combinations, as shown in table~\ref{tab:data_num}.

\begin{table}
\centering
\begin{tabular}{ccccccccc}
\multirow{3}{*}{Combination}
& \multicolumn{8}{c}{Number of data} \\
& \multicolumn{3}{c}{Training--validation} && \multicolumn{3}{c}{Testing} & Total \\
& Splashing& Non-splashing& Total&& Splashing& Non-splashing& Total& \\
1& 114& 87& 201&& 27& 21& 48& 249\\ 
2& 112& 86& 198&& 29& 22& 51& 249\\ 
3& 113& 85& 198&& 28& 23& 51& 249\\ 
4& 114& 85& 199&& 27& 23& 50& 249\\ 
5& 111& 89& 200&& 30& 19& 49& 249\\
\end{tabular}
\caption{
\label{tab:data_num} 
Numbers of splashing and non-splashing data for training--validation and testing in each data combination.
}
\end{table}

\subsection{\label{sec:ann} Feedforward neural network}
\begin{figure}
\centering
\includegraphics[width=\textwidth]{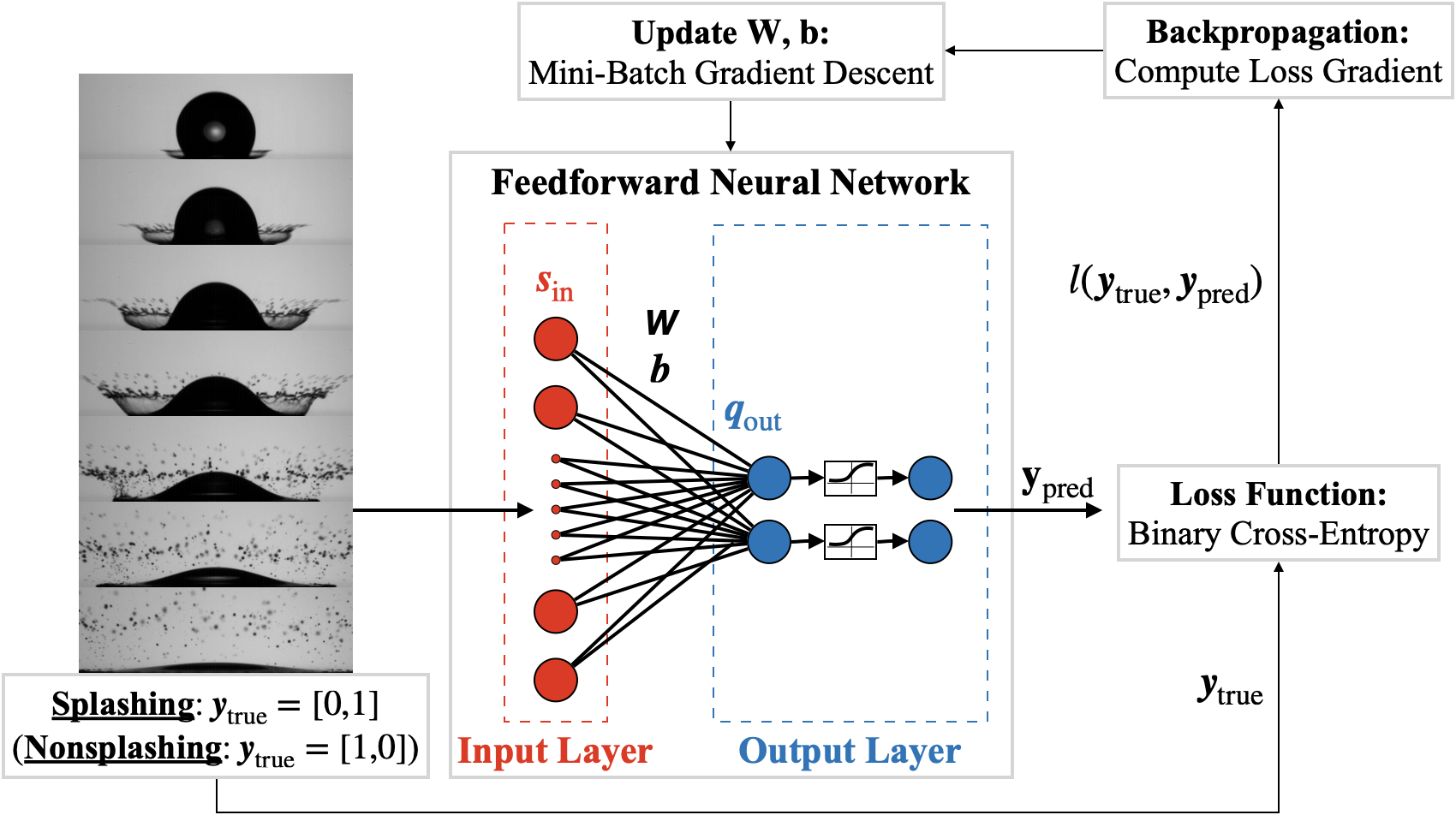}
\caption{\label{fig:fnn_arch_train} 
Training and architecture of the FNN that was used to extract the morphological features of splashing and non-splashing drops.}
\end{figure}

Figure~\ref{fig:fnn_arch_train} illustrates the training and architecture of the FNN developed for the extraction of the critical impact time and the morphological features through the classification of splashing and non-splashing drops based on the image sequences showing the temporal evolution of the drop morphology.
The FNN was implemented in the Python programming language on Jupyter Notebook \citep{kluyver2016jupyter} using the libraries of TensorFlow~\citep{abadi2016tensorflow}. 
The code is available on \href{https://github.com/yeejingzuTUAT/ImageAndImageSequenceClassificationForSplashingAndNonsplashingDrops}{GitHub} (\url{https://github.com/yeejingzuTUAT/ImageAndImageSequenceClassificationForSplashingAndNonsplashingDrops}).

In the input layer, the input image sequence is flattened into a one-dimensional column vector $\boldsymbol{s}_\mathrm{in} \in \mathbb{R}^{M}$, for $M = N_\mathrm{img} h_\mathrm{img} w_\mathrm{img}$, where $N_\mathrm{img}$ is the total number of frames in an image sequence, $h_\mathrm{img}$ is the height of an image in pixels, and $w_\mathrm{img}$ is the width of an image in pixels.
In this study, the values of $h_\mathrm{img}$ and $w_\mathrm{img}$ are 200 and 640, respectively.

Each element of $\boldsymbol{s}_\mathrm{in}$ in the input layer (red circles in figure~\ref{fig:fnn_arch_train}) is fully connected to each element of $\boldsymbol{q}_\mathrm{out}$ in the output layer (blue circles) by a linear function:
\begin{equation}
\boldsymbol{q}_\mathrm{out} = \mathsfbi{W}\boldsymbol{s}_\mathrm{in} + \boldsymbol{b}
\label{eq:lin_func_out},
\end{equation}
where $\boldsymbol{q}_\mathrm{out} \in \mathbb{R}^{C}$ is the output vector, which can be interpreted as a vector containing the prediction values, $\mathsfbi{W} \in \mathbb{R}^{C \times M}$ is the weight matrix, and $\boldsymbol{b} \in \mathbb{R}^{C}$ is the bias vector.
Note that bold italic symbols like $\boldsymbol{s}$ indicate vectors and bold sloping sans serif symbols like $\mathsfbi{W}$ indicate matrices.
$C$ is the total number of classes for classification, which are splashing and non-splashing in this case, and so $C=2$.
The value of each element in $\mathsfbi{W}$ and $\boldsymbol{b}$, which is initialized using the Glorot uniform initializer~\citep{glorot2010understanding}, is determined through the training.

In the output layer, each element of $\boldsymbol{q}_\mathrm{out}$ is activated by a sigmoid function, which saturates negative values at 0 and positive values at 1, as follows:
\begin{equation}
y_{\mathrm{pred},i} = 
\frac{1}{1+\mathrm{e}^{-q_{\mathrm{out},i}}}
\label{eq:sigmoid},
\end{equation}
for $i = 1,\dots,C$, where the activated value ${y}_{\mathrm{pred},i}$ is an element of $\boldsymbol{y}_\mathrm{pred} \in \mathbb{R}^{C}$.
$\boldsymbol{y}_\mathrm{pred} = [y_{\mathrm{pred},1}, y_{\mathrm{pred},2}]$ can be interpreted as a vector containing the probabilities $y_{\mathrm{pred},1}$ and $y_{\mathrm{pred},2}$ of an input image sequence to be classified as a non-splashing drop and as a splashing drop, respectively.
Throughout this paper, the subscripts `$\mathrm{nonspl}$' and `$\mathrm{spl}$' are used instead of the subscripts `$1$' and `$2$', respectively.
Thus, $\boldsymbol{y}_\mathrm{pred} = [y_{\mathrm{pred},1}, y_{\mathrm{pred},2}] = [y_{\mathrm{pred,nonspl}}, y_{\mathrm{pred,spl}}]$.
For training, $\boldsymbol{y}_\mathrm{pred}$ is computed for all training image sequences and compared with the respective true labels $\boldsymbol{y}_\mathrm{true}\in \mathbb{R}^{C}$.
The true labels for the image sequences of a splashing drop and a non-splashing drop are $\boldsymbol{y}_\mathrm{true} = [0,1]$ and $[1,0]$, respectively.

A binary cross-entropy loss function is used for the comparison between $\boldsymbol{y}_\mathrm{pred}$ and $\boldsymbol{y}_\mathrm{true}$ as follows:
\begin{equation}
l(\boldsymbol{y}_\mathrm{true},\boldsymbol{y}_\mathrm{pred})
=\sum_{i=1}^{C}[
-{y}_{\mathrm{true},i}\ln({ y}_{\mathrm{pred},i})-(1-{y}_{\mathrm{true},i})\ln(1-{y}_{\mathrm{pred},i})]
\label{eq:cross_entropy},
\end{equation}
for $i = 1, \dots, C$, where $l$ is the computed loss.
From this equation, the value of $l$ approaches 0 as $\boldsymbol{y}_\mathrm{pred}$ approaches $\boldsymbol{y}_\mathrm{true}$ and increases significantly as $\boldsymbol{y}_\mathrm{pred}$ varies away from $\boldsymbol{y}_\mathrm{true}$.
$l$ is computed during training and validation, but not during testing.

Through a backpropagation algorithm~\citep{rumelhart1986learning}, the gradient of $l$ with respect to each element of $\mathsfbi{W}$ and $\boldsymbol{b}$ of the FNN is computed.
The computed gradient determines whether the value of an element should be increased or decreased, and the amount by which this should be done,
when $\mathsfbi{W}$ and $\boldsymbol{b}$ are updated using the mini-batch gradient descent algorithm~\citep{li2014efficient}.
Regularization of early stopping~\citep{prechelt1998early} is applied to determine when to stop updating $\mathsfbi{W}$ and $\boldsymbol{b}$.

The percentage accuracy of the trained FNN is also evaluated as follows:
\begin{equation}
\text{accuracy}= \frac{\text{number of correct predictions}}{\text{total number of predictions}} \times 100
\label{eq:acc}.
 \end{equation}
The number of correct predictions is determined by the classification threshold.
The trained FNN classifies an image sequence based on the element of $\boldsymbol{y}_\mathrm{pred}$ that has a value equal to or greater than that of the classification threshold.
In this study, the classification threshold is fixed at $0.5$.
For example, if the prediction of an image sequence by the trained FNN is $\boldsymbol{y}_\mathrm{pred} = [0.25,0.75]$, then the image sequence will be classified as an image sequence of a splashing drop.
Accuracy is computed during training, validation, and testing.
\begin{figure}
\centering
\subfloat[]{
\includegraphics[width=0.45\columnwidth]{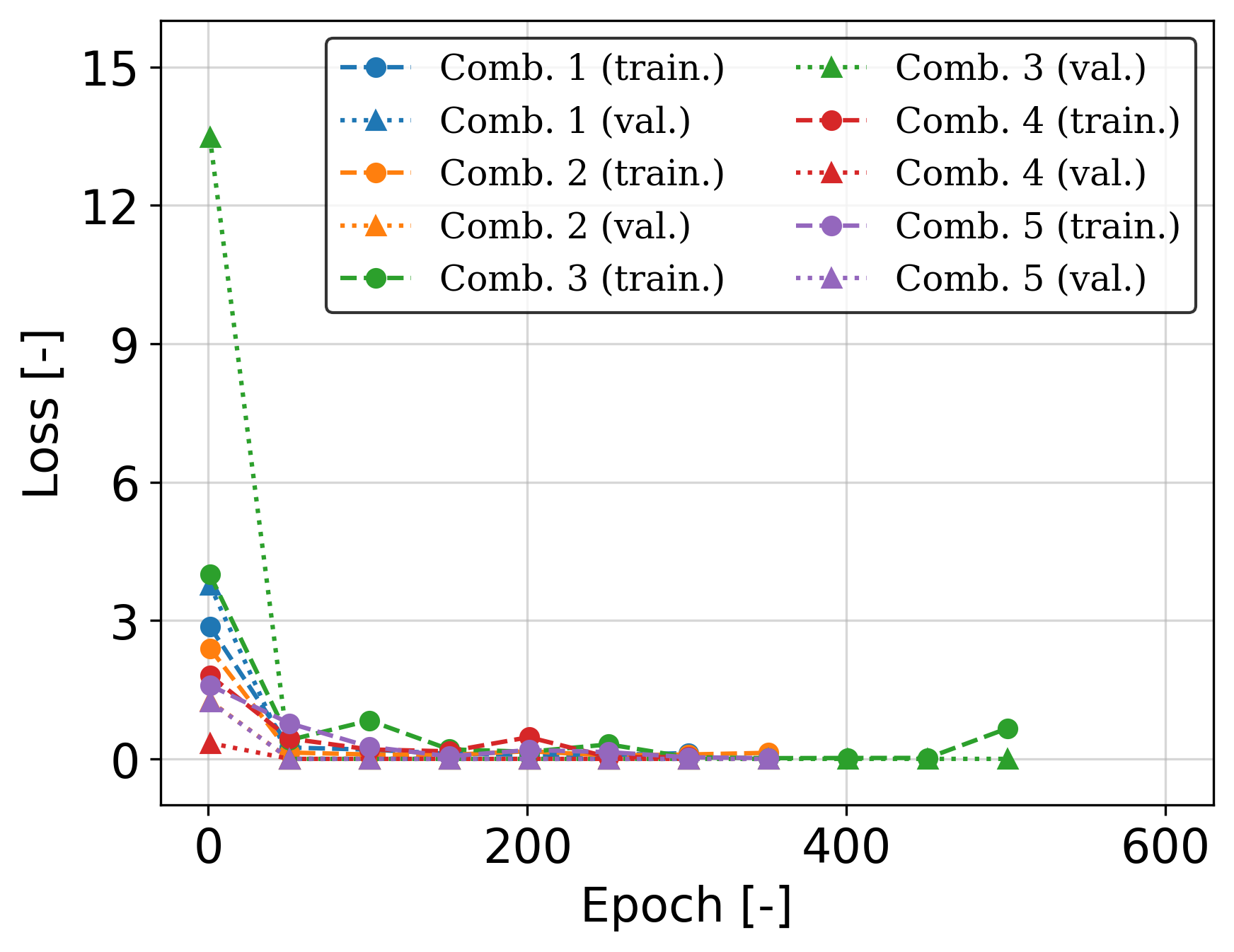}
\label{fig:loss} }
\subfloat[]{
\includegraphics[width=0.45\columnwidth]{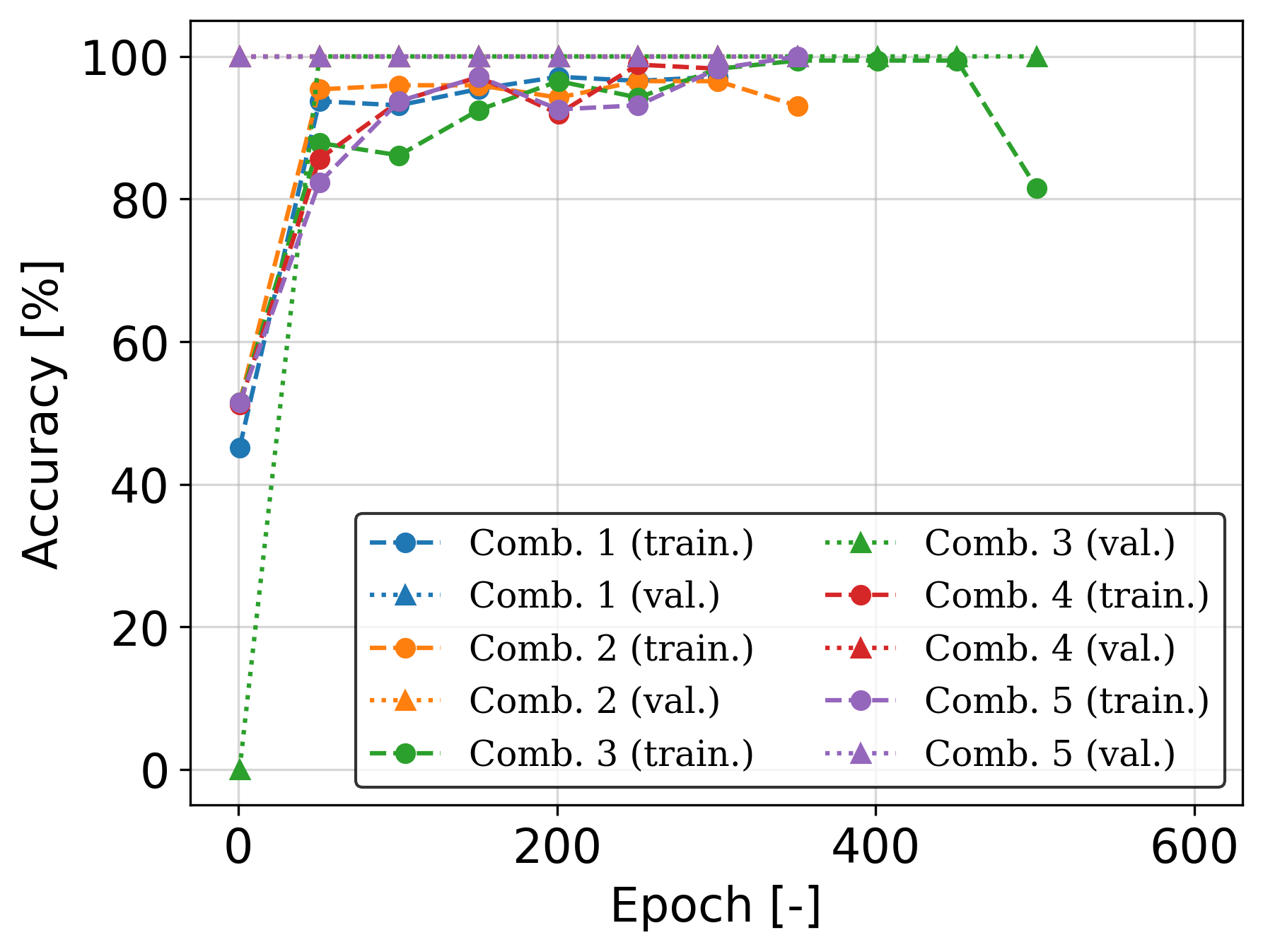}
\label{fig:acc} }
\caption{\label{fig:class_perform}
Training and validation of the FNN for image-sequence classification: (\textit{a}) losses and (\textit{b}) accuracies, averaged among every 50 epochs. 
Comb., combination; train., training; val. validation.}
\end{figure}

The training--validation of the FNN for image-sequence classification was evaluated from the plots of losses and accuracies averaged among every 50 epochs, which are shown in figure~\ref{fig:class_perform}.
Here, the number of epochs indicates how many times all training--validation image sequences were fed through the FNN for training.
As the number of epochs increases, losses decrease and approach 0, while accuracies increase and approach 1.
Early stopping prevents overfitting by stopping the updating of $\mathsfbi{W}$ and $\boldsymbol{b}$ when the losses reach their minimum values.
These trends confirm that the training and validation have been carried out properly and the trained FNN has achieved the desired classification performance.
The trained FNN is then used to classify test image sequences to check their generalizability.

\section{\label{sec:result}Results and discussion}

In \S~\ref{sec:class_perform}, the testing of the trained FNN is explained.
In \S~\ref{sec:class_process}, the analysis of the classification process is discussed.
In \S~\ref{sec:fea_ext}, the process for extracting the features used by the FNN to classify splashing and non-splashing drops is elaborated.
In \S~\ref{sec:crit_time}, the importance index for quantifying the contributions of the extracted features in each frame of an image sequence is introduced and discussed.

\subsection{\label{sec:class_perform}Testing of feedforward neural network}

\begin{figure}
\centering
\subfloat[]{
\includegraphics[width=0.45\columnwidth]{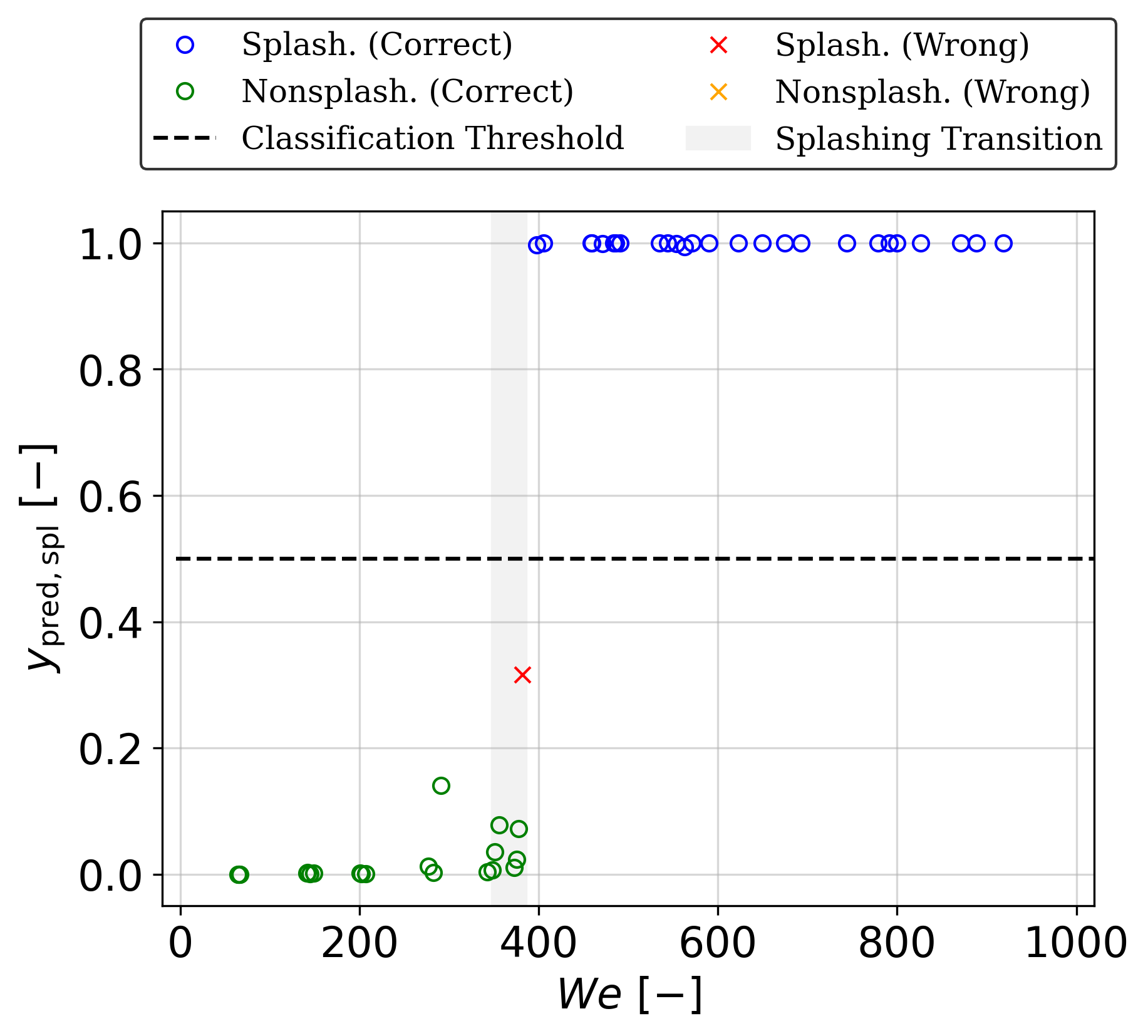}
\label{fig:prob_spl} }
\subfloat[]{
\includegraphics[width=0.45\columnwidth]{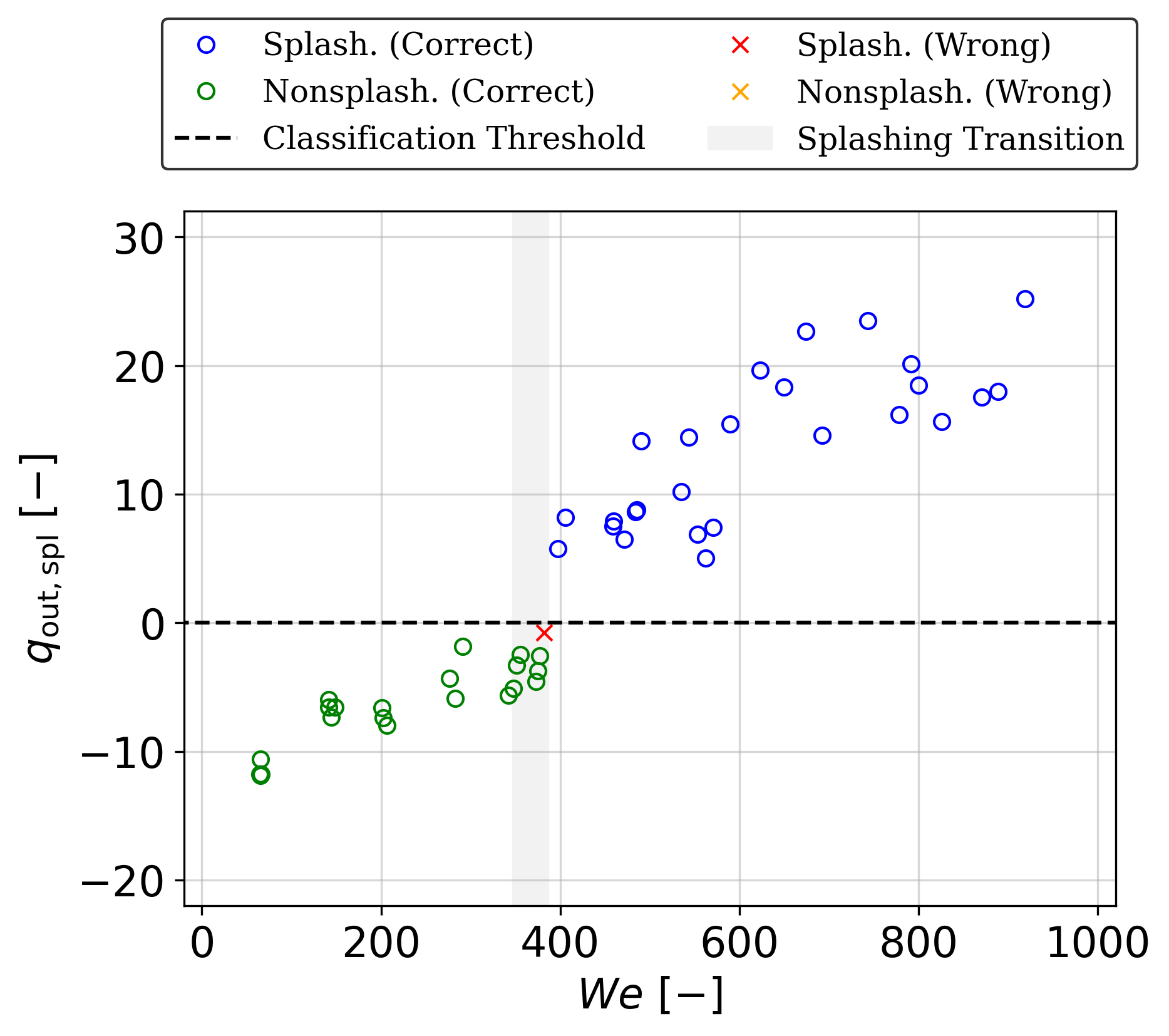}
\label{fig:qout_spl}}
\caption{\label{fig:prob_q_out_spl}
(\textit{a}) Splashing probability $y_{\mathrm{pred,spl}}$ and (\textit{b}) splashing prediction value $q_{\mathrm{out,spl}}$ versus Weber number $\We$ for test image sequences of combination 1.
}
\end{figure}

Testing is the evaluation of the ability of a trained FNN to classify image sequences that were not used to train the FNN.
The results for all data combinations are shown in table~\ref{tab:test_result_img_seq}.
Among all combinations, the test accuracy in classifying image sequences of splashing and non-splashing drops is higher than $96\%$.
The confidence of the classifications performed by the trained FNN can be analysed from the plot of the splashing probability $y_{\mathrm{pred,spl}}$ computed by the trained FNN for the test image sequences.
Since similar results were obtained for all data combinations, only the plot for combination 1 is shown in figure~\ref{fig:prob_spl}.
For most image sequences of splashing and non-splashing drops, the values of $y_{\mathrm{pred,spl}}$ computed by the trained FNN are $\geq$0.8 and $\leq$0.2, respectively.
In other words, most of the computed $y_{\mathrm{pred,spl}}$ differ by at least 0.3 from the classification threshold, which is fixed at 0.5, indicating relatively high confidence of the classifications performed by the trained FNN.
\begin{table}
\centering
\begin{tabular}{ccccccccc}
\multirow{2}{*}{Combination}&
\multicolumn{8}{c}{Test accuracy} \\ &
\multicolumn{2}{c}{Splashing} && \multicolumn{2}{c}{Non-splashing} && \multicolumn{2}{c}{Total} \\
1& 26/27& 96\%&& 21/21& 100\%&& 47/48& 98\% \\
2& 29/29& 100\%&& 22/22& 100\%&& 51/51& 100\% \\
3& 27/28& 96\%&& 22/23& 96\%&& 49/51& 96\% \\
4& 26/27& 96\%&& 23/23& 100\%&& 49/50& 98\% \\
5& 28/30& 93\%&& 19/19& 100\%&& 47/49& 96\%
\end{tabular}
\caption{\label{tab:test_result_img_seq}
Test accuracy of FNN trained with different data combinations in classifying image sequences of splashing and non-splashing drops.
}
\end{table}

The high accuracy and confidence in image-sequence classification by the simple but highly explainable FNN architecture is possible because of the high similarity of the image sequences.
As mentioned in \S~\ref{sec:method}, the drop size and the spatial resolution of the frames in each image sequence were kept constant, with a low standard deviation.
Moreover, the background of the image data was trimmed to ensure that the impacting drop was positioned at the centre of each frame.
As can be seen in figure~\ref{fig:processed_img}, regardless of $H$, $\We$, and the outcome of the impact, the image sequences are very similar but still retain the important distinguishing characteristics of splashing and non-splashing drops.
The classification process of the well-trained FNN can now be visualized and analysed.

The splashing probability $y_{\mathrm{pred,spl}}$ is calculated from the splashing prediction value $q_{\mathrm{out,spl}}$ using a sigmoid function (see~\eqref{eq:sigmoid}), where $y_{\mathrm{pred,spl}} = 0.5$ when $q_{\mathrm{out,spl}} = 0$.
In other words, the trained FNN classifies an image sequence based on $q_{\mathrm{out,spl}} = 0$, where an image sequence is classified as a splashing drop if $q_{\mathrm{out,spl}} \geq 0$ and as a non-splashing drop if $q_{\mathrm{out,spl}} < 0$.
Since $q_{\mathrm{out,spl}}$ is not saturated to between 0 and 1 like $y_{\mathrm{pred,spl}}$, it has a linear relationship with $\We$, as shown in figure~\ref{fig:qout_spl}. 
Such a linear relationship indicates the potential for measuring physical quantities from the image sequence using the FNN.

\subsection{\label{sec:class_process} Analysis of the classification process}

To extract the features of the morphological evolution of splashing and non-splashing drops during the impact and the importance index of the extracted features, it is necessary to analyse the classification process of the FNN.
In~\eqref{eq:lin_func_out}, the two elements of the prediction values vector $\boldsymbol{q}_{\mathrm{out}}$ are the splashing prediction value $q_{\mathrm{out, spl}}$ and the non-splashing prediction value $q_{\mathrm{out, nonspl}}$.
Also, the weight matrix $\mathsfbi{W}$ can be decomposed into two row vectors $\boldsymbol{w}_{\mathrm{spl}}$ and $\boldsymbol{w}_{\mathrm{nonspl}}$, while the bias vector $\boldsymbol{b}$ can be decomposed into two elements $b_{\mathrm{spl}}$ and $b_{\mathrm{nonspl}}$.
Hence, the elaborated form of~\eqref{eq:lin_func_out} can be expressed as
\begin{equation}
\boldsymbol{q}_\mathrm{out} = 
\begin{bmatrix}
q_{\mathrm{out, nonspl}}\\
q_{\mathrm{out, spl}}
\end{bmatrix}
=
\begin{bmatrix}
\boldsymbol{w}_{\mathrm{nonspl}} \boldsymbol{\cdot} \boldsymbol{s}_{\mathrm{in}}\\
\boldsymbol{w}_{\mathrm{spl}} \boldsymbol{\cdot} \boldsymbol{s}_{\mathrm{in}}
\end{bmatrix}
+
\begin{bmatrix}
b_{\mathrm{nonspl}}\\b_{\mathrm{spl}}
\end{bmatrix}
\approx
\begin{bmatrix}
\boldsymbol{w}_{\mathrm{nonspl}} \boldsymbol{\cdot} \boldsymbol{s}_{\mathrm{in}}\\
\boldsymbol{w}_{\mathrm{spl}} \boldsymbol{\cdot} \boldsymbol{s}_{\mathrm{in}}
\end{bmatrix}
\label{eq:lin_func_out_dissect},
\end{equation}
where $\boldsymbol{s}_{\mathrm{in}}$ is an image sequence flattened into a vector.
The products $\boldsymbol{w}_{\mathrm{spl}} \boldsymbol{\cdot} \boldsymbol{s}_{\mathrm{in}}$ and $\boldsymbol{w}_{\mathrm{nonspl}} \boldsymbol{\cdot} \boldsymbol{s}_{\mathrm{in}}$ are given by
\begin{align}
\boldsymbol{w}_{\mathrm{spl}} \boldsymbol{\cdot} \boldsymbol{s}_{\mathrm{in}}
&= w_{\mathrm{spl},1}s_{\mathrm{in},1} + w_{\mathrm{spl},2}s_{\mathrm{in},2} + \dots + w_{\mathrm{spl},M}s_{\mathrm{in},M}
= \sum_{i = 1}^{M} w_{\mathrm{spl},i}s_{\mathrm{in},i}
\label{eq:q_out_spl},
\\
\boldsymbol{w}_{\mathrm{nonspl}} \boldsymbol{\cdot} \boldsymbol{s}_{\mathrm{in}}
&= w_{\mathrm{nonspl},1}s_{\mathrm{in},1} + w_{\mathrm{nonspl},2}s_{\mathrm{in},2} + \dots + w_{\mathrm{nonspl},M}s_{\mathrm{in},M}
= \sum_{i = 1}^{M} w_{\mathrm{nonspl},i}s_{\mathrm{in},i}
\label{eq:q_out_nonspl},
 \end{align}
respectively, for $M = N_\mathrm{img} h_\mathrm{img} w_\mathrm{img}$.
Here, $\boldsymbol{q}_\mathrm{out} \approx \mathsfbi{W}\boldsymbol{s}_{\mathrm{in}}$, because the elements of the trained $\boldsymbol{b}$ are negligible.
Among the test image sequences of all data combinations, the element of $\boldsymbol{q}_{\mathrm{out}}$ with the smallest absolute value computed by the trained FNN is 0.0327.
On the other hand, the elements of $\boldsymbol{b}$ are of the order of $10^{-4}$, which is much smaller than the element of $\boldsymbol{q}_{\mathrm{out}}$ with the smallest absolute value.

Despite $\boldsymbol{q}_{\mathrm{out}}$ having two elements, an analysis based on either element is sufficient, because the absolute values of the two elements are approximately equal.
In binary classification, the probabilities of both output classes should sum to one, which in this case means that $y_{\mathrm{pred, spl}} + y_{\mathrm{pred, nonspl}} \approx 1$, where 
 the approximately equals sign is used to compensate for the truncation error.
Owing to the use of the sigmoid function (see~\eqref{eq:sigmoid}), when $y_{\mathrm{pred, spl}} + y_{\mathrm{pred, nonspl}} \approx 1$, the values of the two elements of $\boldsymbol{q}_{\mathrm{out}}$ are approximately equal: $q_{\mathrm{out, spl}} \approx - q_{\mathrm{out, nonspl}}$.

For the analysis of $q_{\mathrm{out, spl}}$, the elements $w_{\mathrm{spl},i}s_{\mathrm{in},i}$ in~\eqref{eq:q_out_spl} are grouped according to each frame:
\begin{align}
\boldsymbol{w}_{\mathrm{spl},z_0/D_0} \boldsymbol{\cdot} \boldsymbol{s}_{\mathrm{in},z_0/D_0}
&= w_{\mathrm{spl},z_0/D_0,1}s_{\mathrm{in},z_0/D_0,1} + \dots + w_{\mathrm{spl},z_0/D_0,m}s_{\mathrm{in},z_0/D_0,m}\nonumber
\\&= \sum_{i = 1}^{m} w_{\mathrm{spl},z_0/D_0,i}s_{\mathrm{in},z_0/D_0,i}
\label{eq:q_out_z0D0_ij},
 \end{align}
where $\boldsymbol{s}_{\mathrm{in},z_0/D_0} \in \mathbb{R}^{m}$ is a frame flattened into a vector, $\boldsymbol{w}_{\mathrm{spl},z_0/D_0} \in \mathbb{R}^{m}$ is the vector that contains the elements of the splashing weight vector $\boldsymbol{w}_{\mathrm{spl}}$ that corresponds to a frame, and $m$ $(= h_\mathrm{img}w_\mathrm{img})$ is the total number of pixels in a frame. 
Thus,~\eqref{eq:q_out_spl} can be expressed as
\begin{align}
\boldsymbol{w}_{\mathrm{spl}} \boldsymbol{\cdot} \boldsymbol{s}_{\mathrm{in}}
&= \boldsymbol{w}_{\mathrm{spl},z_0/D_0 = 0.875} \boldsymbol{\cdot} \boldsymbol{s}_{\mathrm{in},z_0/D_0 = 0.875} + \dots + \boldsymbol{w}_{\mathrm{spl},z_0/D_0 = 0.125} \boldsymbol{\cdot} \boldsymbol{s}_{\mathrm{in},z_0/D_0 = 0125}\nonumber
\\&= \sum{\boldsymbol{w}_{\mathrm{spl},z_0/D_0} \boldsymbol{\cdot} \boldsymbol{s}_{\mathrm{in},z_0/D_0}}
\label{eq:q_out_spl_z0D0}.
 \end{align}
Since $q_{\mathrm{out, spl}}\approx \boldsymbol{w}_{\mathrm{spl}} \boldsymbol{\cdot} \boldsymbol{s}_{\mathrm{in}}$, the value of $\boldsymbol{w}_{\mathrm{spl},z_0/D_0} \boldsymbol{\cdot} \boldsymbol{s}_{\mathrm{in},z_0/D_0}$ for each frame shows the respective contribution to the computation of $q_{\mathrm{out,spl}}$.
The morphological features and the importance index can be extracted using $\boldsymbol{w}_{\mathrm{spl},z_0/D_0}$ and $\boldsymbol{w}_{\mathrm{spl},z_0/D_0} \boldsymbol{\cdot} \boldsymbol{s}_{\mathrm{in},z_0/D_0}$, respectively.

\subsection{\label{sec:fea_ext} Extraction of morphological features}

In this subsection, the extraction of the features of the morphological evolution of splashing and non-splashing drops is explained.
As mentioned in \S~\ref{sec:class_perform}, the trained FNN classifies an image sequence based on $q_{\mathrm{out, spl}} = 0$, where an image sequence is classified as a splashing drop if $q_{\mathrm{out,spl}} \geq 0$ and as a non-splashing drop if $q_{\mathrm{out,spl}} < 0$.
Since $q_{\mathrm{out,spl}} \approx \boldsymbol{w}_{\mathrm{spl}} \boldsymbol{\cdot} \boldsymbol{s}_{\mathrm{in}} = \sum{\boldsymbol{w}_{\mathrm{spl},z_0/D_0} \boldsymbol{\cdot} \boldsymbol{s}_{\mathrm{in},z_0/D_0}}$, the value of $\boldsymbol{w}_{\mathrm{spl},z_0/D_0} \boldsymbol{\cdot} \boldsymbol{s}_{\mathrm{in},z_0/D_0}$ of each frame has to be as high as possible for a splashing drop to have $q_{\mathrm{out,spl}} \geq 0$.
On the other hand, the value of $\boldsymbol{w}_{\mathrm{spl},z_0/D_0} \boldsymbol{\cdot} \boldsymbol{s}_{\mathrm{in},z_0/D_0}$ of each frame has to be as low as possible for a non-splashing drop to have $q_{\mathrm{out,spl}} < 0$.

\begin{figure}
\centering
\includegraphics[width=\textwidth]{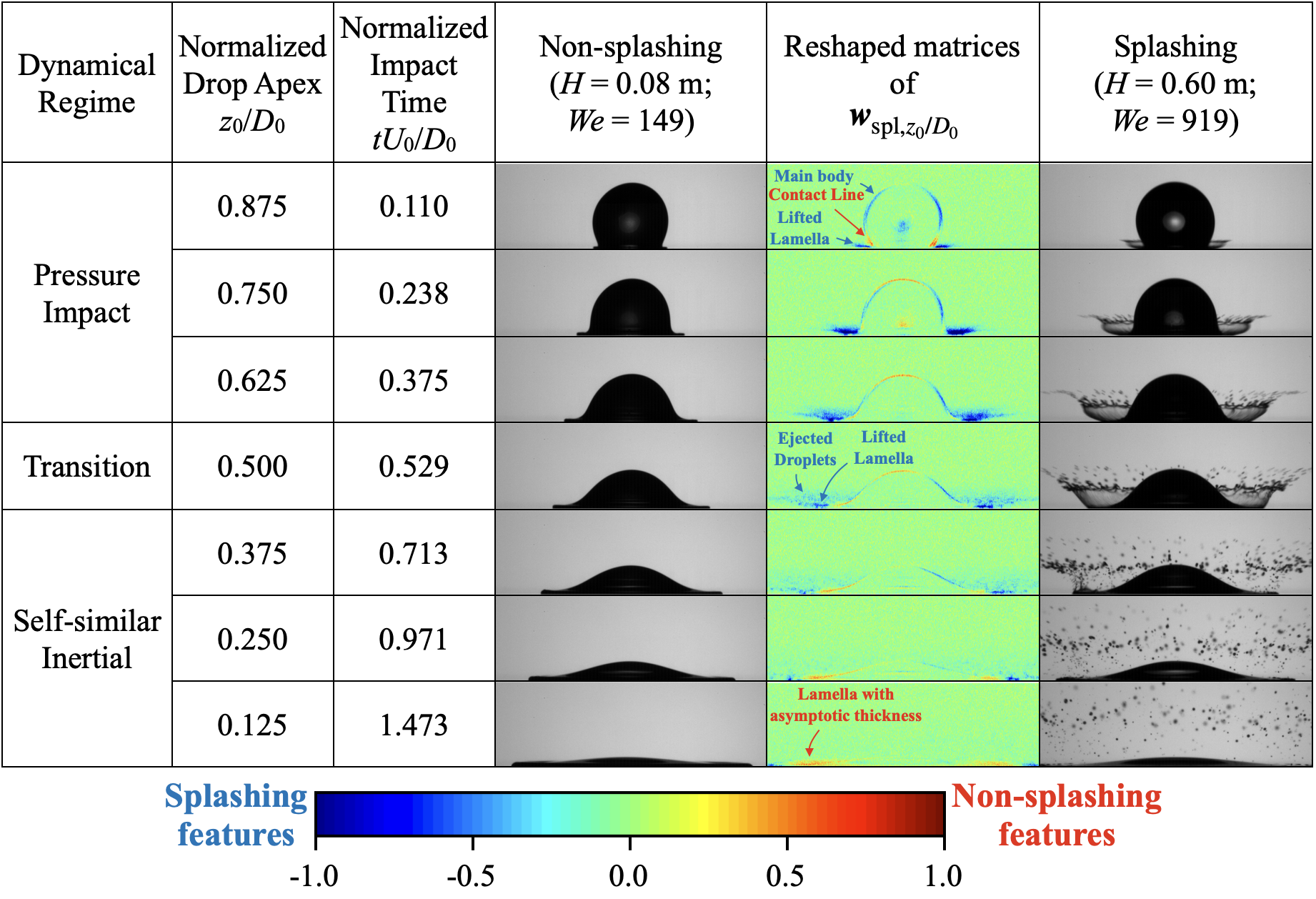}
\caption{
\label{fig:trained_w}
Colour maps of the reshaped matrices of ${\boldsymbol w}_{\mathrm{spl},z_0/D_0}$ of the FNN when trained with combination 1.
The distributions were similar for the FNN when it was trained with other combinations.
}
\end{figure}

For the analysis of $\boldsymbol{w}_{\mathrm{spl},z_0/D_0} \boldsymbol{\cdot} \boldsymbol{s}_{\mathrm{in},z_0/D_0}$, the $\boldsymbol{w}_{\mathrm{spl},z_0/D_0}$ vector of each frame is reshaped in row-major order into a two-dimensional $h_\mathrm{img} \times w_\mathrm{img}$ matrix $\mathsfbi{w}_{\mathrm{spl},z_0/D_0}$, which is the shape of a frame: $\boldsymbol{w}_{\mathrm{spl},z_0/D_0} \in \mathbb{R}^{m} \to \mathsfbi{w}_{\mathrm{spl},z_0/D_0} \in \mathbb{R}^{h_\mathrm{img} \times w_\mathrm{img}}$.
The reshaped matrices  $\mathsfbi{w}_{\mathrm{spl},z_0/D_0}$ are visualized as colour maps.
For explanation, the colour maps of the reshaped matrices of $\boldsymbol{w}_{\mathrm{spl},z_0/D_0}$ of the FNN trained with combination 1 are presented in figure~\ref{fig:trained_w}.
The values of the elements in ${\boldsymbol w}_{\mathrm{spl},z_0/D_0}$ are normalized by the maximum absolute values in ${\boldsymbol w}_{\mathrm{spl}}$, and thus the blue--green--red (BGR) scale is from $-1.0$ to 1.0.
Note that only the colour maps of combination 1 are shown, because those for the other combinations are similar.

In the colour maps, the distributions of the extreme values, i.e., values with large magnitudes, show the important features that the FNN identifies to classify splashing and non-splashing drops.
The extreme negative values shown in blue are the features of splashing drops, and the extreme positive values shown in red are the features of non-splashing drops.
This is because the high-speed videos were captured using shadowgraphy, where the normalized intensity value of a pixel occupied by the drop is zero ($s_{\mathrm{in},z_0/D_0,i} = 0$), while the normalized intensity value of a pixel not occupied by the drop but capturing the backlight is near to one ($s_{\mathrm{in},z_0/D_0,i} \to 1$).
Hence, the FNN assigned negative values to those pixel positions occupied by a splashing drop but not by a non-splashing drop, and so those negative values would be cancelled out by a splashing drop ($w_{\mathrm{in},z_0/D_0,i}s_{\mathrm{in},z_0/D_0,i} = 0$) to increase the value of $\boldsymbol{w}_{\mathrm{spl},z_0/D_0} \boldsymbol{\cdot} \boldsymbol{s}_{\mathrm{in},z_0/D_0}$.
However, for a non-splashing drop, those negative values would not be cancelled out and would remain ($w_{\mathrm{in},z_0/D_0,i}s_{\mathrm{in},z_0/D_0,i} \to w_{\mathrm{in},z_0/D_0,i}$), thereby reducing the value of $\boldsymbol{w}_{\mathrm{spl},z_0/D_0} \boldsymbol{\cdot} \boldsymbol{s}_{\mathrm{in},z_0/D_0}$.
On the other hand, the FNN assigned positive values to those pixel positions occupied by a non-splashing drop but not by a splashing drop, and so those positive values would be cancelled out by a non-splashing drop ($w_{\mathrm{in},z_0/D_0,i}s_{\mathrm{in},z_0/D_0,i} = 0$), reducing the value of $\boldsymbol{w}_{\mathrm{spl},z_0/D_0} \boldsymbol{\cdot} \boldsymbol{s}_{\mathrm{in},z_0/D_0}$.
However, for a splashing drop, those positive values would not be cancelled out and would remain ($w_{\mathrm{in},z_0/D_0,i}s_{\mathrm{in},z_0/D_0,i} \to w_{\mathrm{in},z_0/D_0,i}$), thereby increasing the value of $\boldsymbol{w}_{\mathrm{spl},z_0/D_0} \boldsymbol{\cdot} \boldsymbol{s}_{\mathrm{in},z_0/D_0}$.

By comparing the distribution of the extreme values in the colour maps of the reshaped matrices of ${\boldsymbol w}_{\mathrm{spl},z_0/D_0}$ with the image sequences of typical splashing and non-splashing drops, it is found that the distribution of the values of large magnitudes, i.e. the splashing and non-splashing features, resembles the morphology of an impacting drop.
The distributions of the splashing and non-splashing features indicate that the main morphological differences are the lamella, the contour of the main body, and the ejected secondary droplets.

In terms of the lamella, that of a splashing drop is ejected faster before lifting and breaking into secondary droplets, while that of a non-splashing drop is ejected more slowly before developing into a thicker film.
The differences in the ejection velocity can be seen from the reshaped matrix of ${\boldsymbol w}_{\mathrm{spl},z_0/D_0 = 0.875}$, where the splashing features are distributed around the ejected lamella while the non-splashing features are distributed around the contact line.
\citet{riboux2017boundary} found that in the limit of Ohnesorge number $\Oh$ much smaller than one, the ejection time of the lamella scales with Weber number as $\We^{-2/3}$.
In this study, $\Oh$ is of the order of $10^{-3}$, which is small enough for the scaling found by \citet{riboux2017boundary} to be valid.
Therefore, owing to the higher $\We$ of a splashing drop, the ejection time of the lamella is shorter than in the case of a non-splashing drop.
Furthermore, \citet{philippi2016drop} reported that the pressure peak is near the contact line, causing a bypass motion of the flow.
As a result of a slower ejection of the lamella of a non-splashing drop, more of the volume of the drop is concentrated near the contact line.

From the distribution of the splashing features in ${\boldsymbol w}_{\mathrm{spl},z_0/D_0}$ of $0.625 \leq z_0/D_0 \leq 0.875$, we can see that the lamella of a splashing drop is lifted higher.
A lifted lamella has been identified as being characteristic of a splashing drop by \citet{riboux2014experiments}, who noted that splashing occurs as a result of the vertical lift force imparted by the air on the lamella.
As $z_0/D_0$ reduces to $z_0/D_0 \leq 0.500$, the lamella of a splashing drop descends, and the ejected secondary droplets are too scattered to be captured easily by the FNN.

The distributions of the non-splashing features in ${\boldsymbol w}_{\mathrm{spl},z_0/D_0}$ for $0.125 \leq z_0/D_0 \leq 0.375$ show that the lamella of a non-splashing drop develops into a film thicker than that of a splashing drop.
This can be explained using the studies by \citet{lagubeau2012spreading} and \citet{eggers2010drop}, who reported that in the viscous plateau regime, the asymptotic film thickness scales with $\Rey^{-2/5}D_0$.
Here, $\Rey= \rho U_0 D_0 / \mu$ is the Reynolds number.
In this study, $D_0$ and $\mu$ are the same for all splashing and non-splashing drops, and thus the non-splashing drops have a thicker film owing to the lower $U_0$.

Splashing features can also be found around the contour of the main body, even when $z_0/D_0 = 0.875$.
This indicates that once the impact has commenced, the contour of the main body of a splashing drop is already higher than that of a non-splashing drop.
Most previous studies reported that during the pressure impact regime, the top half of the drop did not yet experience the effects of the impact and kept moving towards the surface at the impact velocity $U_0$, together with the drop apex~\citep{eggers2010drop, gordillo2018dynamics, mitchell2019transient}.
Nevertheless, such small differences in the contour of the main body between splashing and non-splashing drops could be captured using the FNN and were first reported by \citet{yee2022image}, who classified  images of splashing and non-splashing drops using an FNN.

\begin{figure}
\centering
\includegraphics[width=\textwidth]{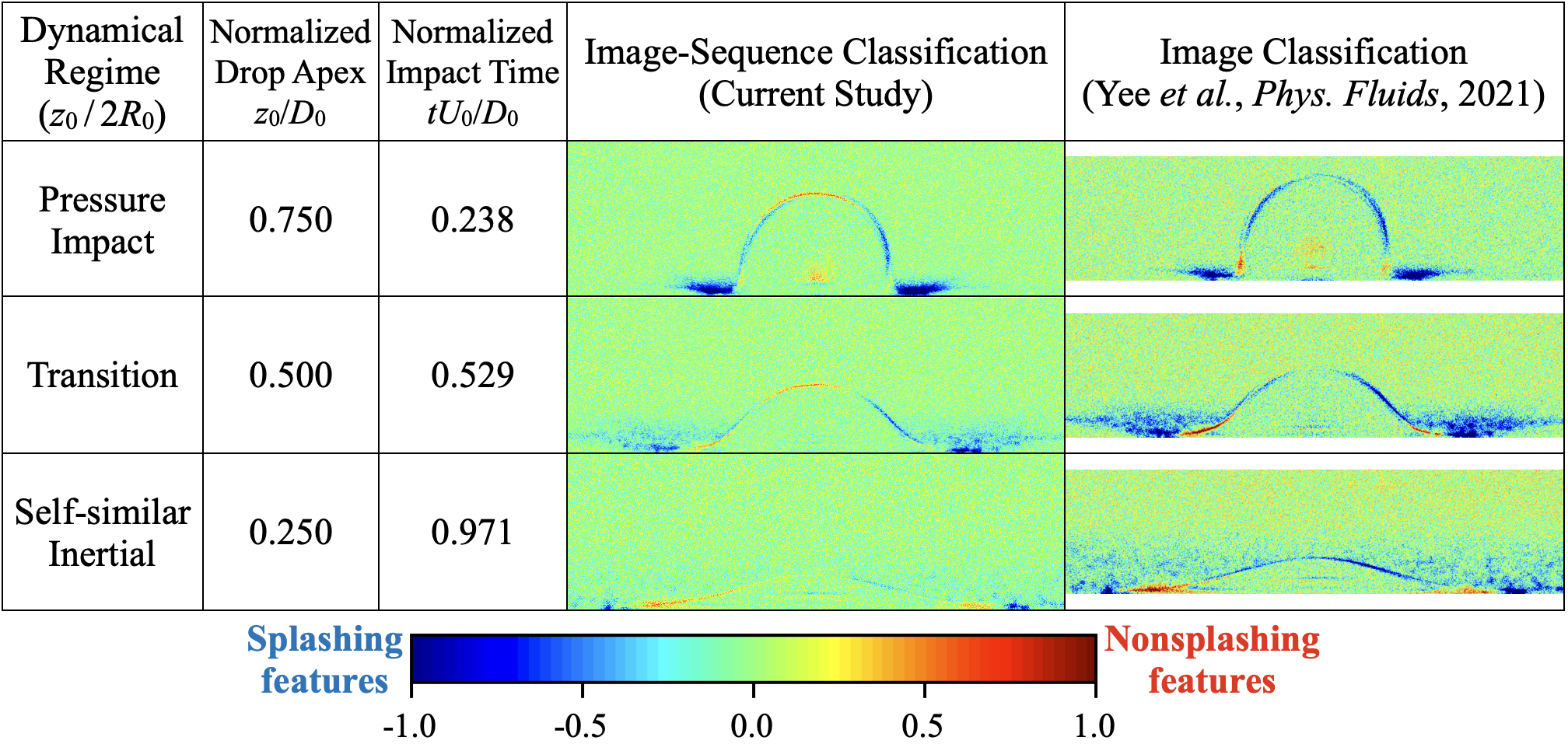}
\caption{
\label{fig:imgSeq_vs_img_class}
Comparison of the colour maps of the reshaped weight vectors trained using image-sequence classification in the current study and image classification in the study by \citet{yee2022image}.
}
\end{figure}

It is important to mention that as $z_0/D_0$ reduces to $z_0/D_0 \leq 0.500$, the distribution of the splashing features around the contour of the main body becomes less obvious.
This is different from the reshaped weight vectors trained using image classification in the study by \citet{yee2022image}.
In their study, they trained three different FNNs to classify splashing and non-splashing drops using  images extracted when $z_0/D_0 = 0.750$, 0.500, and 0.250, respectively.
As shown in figure~\ref{fig:imgSeq_vs_img_class}, the distributions of the splashing and non-splashing features in the reshaped weight vectors trained by 
\citet{yee2022image} are found at similar pixel positions to those in the current study.
Note that in this study, each frame in the image sequences was cropped to $200 \times 640$\,px$^2$, whereas \citet{yee2022image} cropped their images to $160 \times 640$\,px$^2$.
However, the distributions are much more obvious for the reshaped vectors trained using only images of $z_0/D_0 = 0.250$ than for ${\boldsymbol w}_{\mathrm{spl},z_0/D_0 = 0.250}$ trained using image sequences.
This is because the FNN image classifier can only extract information from a single image, whereas the FNN image-sequence classifier of this study can extract information from seven frames in an image sequence.
In other words, the FNN image-sequence classifier can pick and choose the frame from which it wants to extract the information.
Although it could possibly miss important morphological features of splashing and non-splashing drops, quantification of the contribution of each frame and the extracted features to the classification by the FNN image-sequence classifier could provide deeper insights into the morphological evolution of splashing and non-splashing drops, which is discussed in the next section.

\subsection{\label{sec:crit_time}Importance index of the extracted features}

\begin{figure}
\centering
\subfloat[]{
\includegraphics[width=0.8\textwidth]{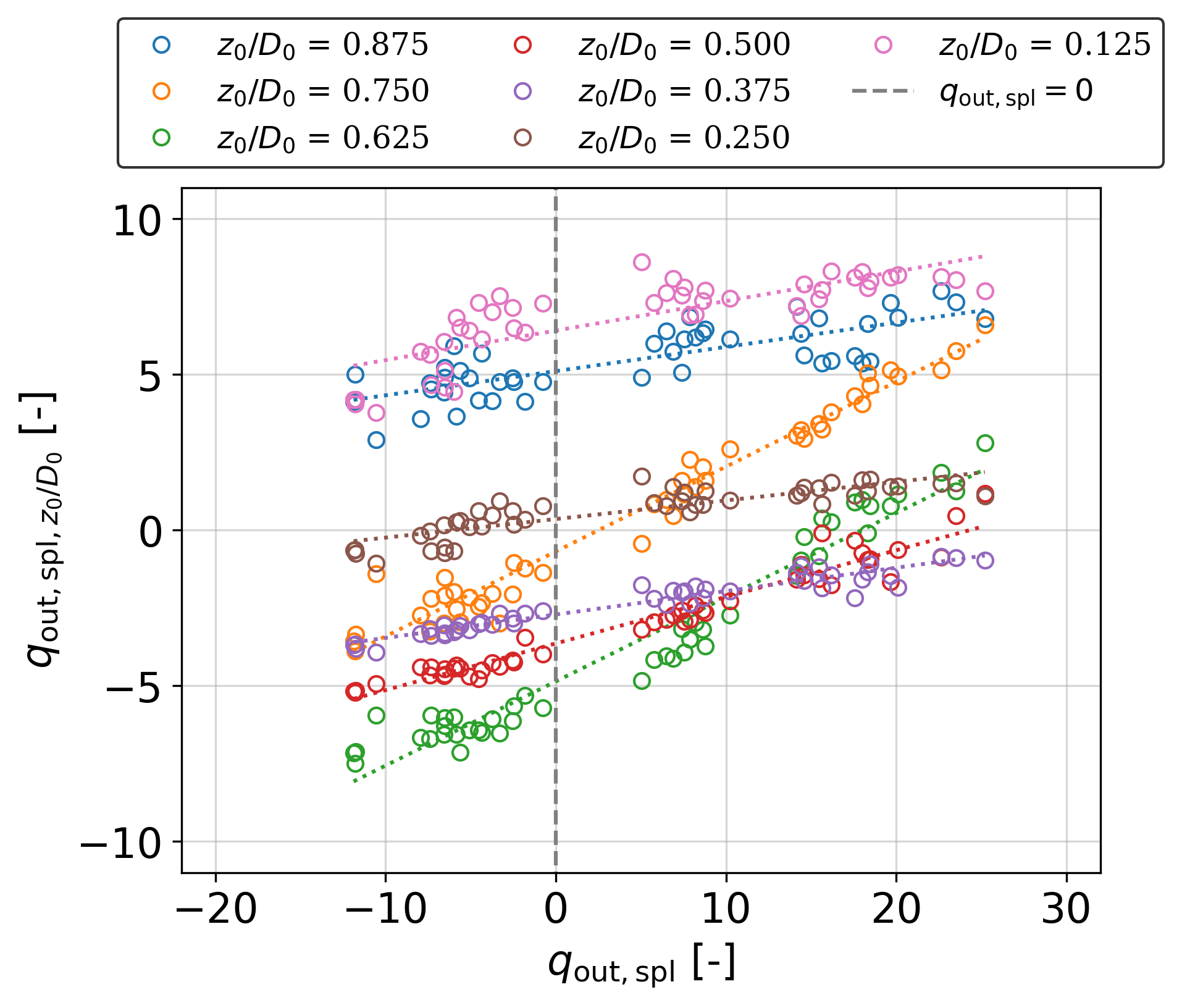}
\label{fig:qout_BP} }
\\
\centering
\subfloat[]{
\includegraphics[width=0.8\columnwidth]{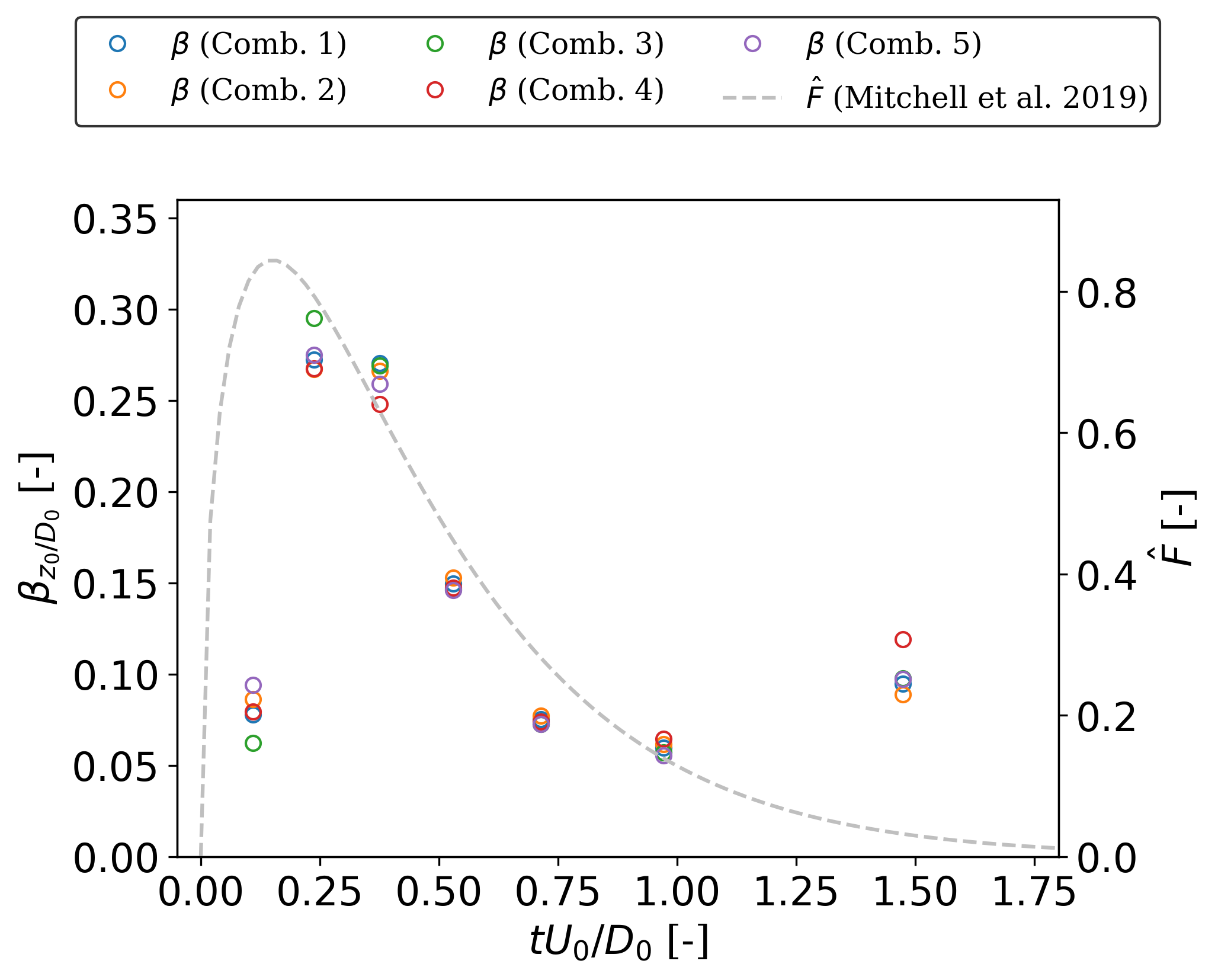}
\label{fig:qGrad}
}
\caption{
\label{fig:qout_BP_qGrad} 
(\textit{a}) $q_{\mathrm{out,spl},z_0/D_0}$ versus $q_{\mathrm{out,spl}}$ of test image sequences of combination 1.
(\textit{b}) Slopes of fitted lines $\beta_{z_0/D_0}$ versus normalized impact time $tU_0/D_0$ of all data combinations.
}
\end{figure}

The importance index for quantifying the contributions of the extracted features in each frame of an image sequence to the classification of the FNN are introduced and discussed in this subsection.
As mentioned in \S~\ref{sec:class_process}, the value of $\boldsymbol{w}_{\mathrm{spl},z_0/D_0} \boldsymbol{\cdot} \boldsymbol{s}_{\mathrm{in},z_0/D_0}$ for each frame shows the respective contribution to the computation of $q_{\mathrm{out,spl}}$.
Denoted by $q_{\mathrm{out,spl},z_0/D_0}$, the values of $\boldsymbol{w}_{\mathrm{spl},z_0/D_0} \boldsymbol{\cdot} \boldsymbol{s}_{\mathrm{in},z_0/D_0}$ for each value of $z_0/D_0$ are plotted against $q_{\mathrm{out,spl}}$.
In figure~\ref{fig:qout_BP}, only the plot for combination 1 is shown, because similar results were obtained for the other data combinations.
The black dashed line shows $q_{\mathrm{out,spl}} = 0$, which corresponds to the classification threshold $y_{\mathrm{pred,spl}} = 0.5$.
To the left of this line where $q_{\mathrm{out,spl}} \geq 0$, an image sequence is classified as that of a splashing drop, while to the right of this line where $q_{\mathrm{out,spl}} < 0$, an image sequence is classified as that of a non-splashing drop.
To identify the importance of each $z_0/D_0$ for the classification of the FNN, least squares fitting is performed for each $z_0/D_0$ and shown in figure~\ref{fig:qout_BP} by the dotted lines with the same colours as the respective markers.
Along with the values of $q_{\mathrm{out,spl}}$, the values of $q_{\mathrm{out,spl},z_0/D_0}$ of all $z_0/D_0$ exhibit an increasing trend, where the slopes of all the fitted lines are positive.
Here, we argue that the $z_0/D_0$ with the slope of the highest value has the most influence on the classification of the FNN.
This is because if the slope has a low value, then the value of $q_{\mathrm{out,spl},z_0/D_0}$ remains constant regardless of the value of $q_{\mathrm{out,spl}}$.
In other words,  $q_{\mathrm{out,spl},z_0/D_0}$ is similar regardless of whether the classification is splashing or non-splashing.
On the contrary, if the slope has a high value, the change in the value of $q_{\mathrm{out,spl},z_0/D_0}$ contributes significantly to the change in the value of $q_{\mathrm{out,spl}}$.
Thus, the classification of an image sequence as that of a splashing or a non-splashing drop is highly dependent on the value of $q_{\mathrm{out,spl},z_0/D_0}$.

Here, the slope of a fitted line is introduced as the importance index for quantifying the contribution of each frame in an image sequence to the classification of the FNN.
Denoted by $\beta_{z_0/D_0}$, the slopes of the fitted lines are plotted against $tU_0/D_0$ for the respective $z_0/D_0$ for all data combinations in figure~\ref{fig:qGrad}.
As can be seen, all data combinations have peak values at $tU_0/D_0 = 0.24$ and 0.38, corresponding to $z_0/D_0 = 0.750$ and 0.625.
These peak values, which range between 0.25 and 0.30, are more than approximately double  the values at other $tU_0/D_0$, which are less than 0.15.
This indicates that the two frames at $tU_0/D_0 = 0.24$ and 0.38 have significantly more influence on the classification by the FNN than the frames at other $tU_0/D_0$.
This is also  why the distributions of the splashing and non-splashing features are  most obvious in the reshaped matrices of $\boldsymbol{w}_{\mathrm{spl},z_0/D_0}$ at $tU_0/D_0 = 0.24$ and 0.38, as shown in figure~\ref{fig:trained_w}.
Therefore, the morphological differences between splashing and non-splashing drops are  most pronounced at $tU_0/D_0 = 0.24$ and 0.38, rather than at the earlier impact time.
These findings are interesting because, for human eyes, splashing drops might look more different from  non-splashing drops at  later impact times.

A question follows from the findings above: why is the morphological difference most pronounced at $tU_0/D_0 = 0.24$ and 0.38?
The answer can be related to the impact force.
Almost all  drop impact studies  focused on the kinematics or the morphology of a drop during  impact, until the recent publication of several studies  focusing on  drop dynamics, specifically the temporal evolution of the impact force and the shear stress distribution beneath impacting drops~\citep{cheng2022drop, philippi2016drop, sun2022stress}.
Specifically, \citet{gordillo2018dynamics} and \citet{mitchell2019transient} reported that although the impact force $F$ increases with increasing $\Rey$, the value of $\hat{F}$, which is $F$ normalized by $\rho U_0^2 D_0^2$, decreases with increasing $\Rey$.
Nevertheless, for inertia-dominated drop impacts, the transient normalized impact force profile can be approximated as
\begin{equation}
\hat{F} 
= \sqrt{\frac{c^2}{\tau}\frac{tU_0}{D_0}}\exp\!\left(-\frac{1}{\tau}\frac{tU_0}{D_0}\right)
\label{eq:F_t},
\end{equation}
where $c$ and $\tau$ are constants, with $c^2/\tau = 1000 \pi /243$ and $1/\tau =10/3$, according to \citet{mitchell2019transient}.
They also reported that the magnitude of the normalized peak impact force is approximately 0.85 and occurs at approximately $tU_0/D_0 = 0.15$ when the side walls of the drop are perpendicular to the impacted surface.
This is consistent with the present study, because careful observation of the examples of the image sequences in figure~\ref{fig:processed_img} shows that the side walls of the drops are perpendicular to the impact surface in the range $0.110 < tU_0/D_0 < 0.238$ when $0.750 < z_0/D_0 < 0.875$.
For comparison with the $\beta_{z_0/D_0}$ values, the line profile of $\hat{F}$ proposed by \citet{mitchell2019transient} is plotted as the grey dashed line in figure~\ref{fig:qGrad}.
As can be seen, the $tU_0/D_0$ of the peak values of $\beta_{z_0/D_0}$ are almost immediately after $tU_0/D_0 = 0.15$, indicating that the morphological differences between splashing and non-splashing drops are most pronounced after the peak normalized impact force $\hat{F}$ has been exerted on the surface.
Since a splashing drop has a higher $U_0$ than a non-splashing drop, the peak non-normalized impact force $F$ exerted by a splashing drop is greater than that exerted by a non-splashing drop.

\begin{figure}
\centering
\subfloat[]{
\includegraphics[width=0.8\textwidth]{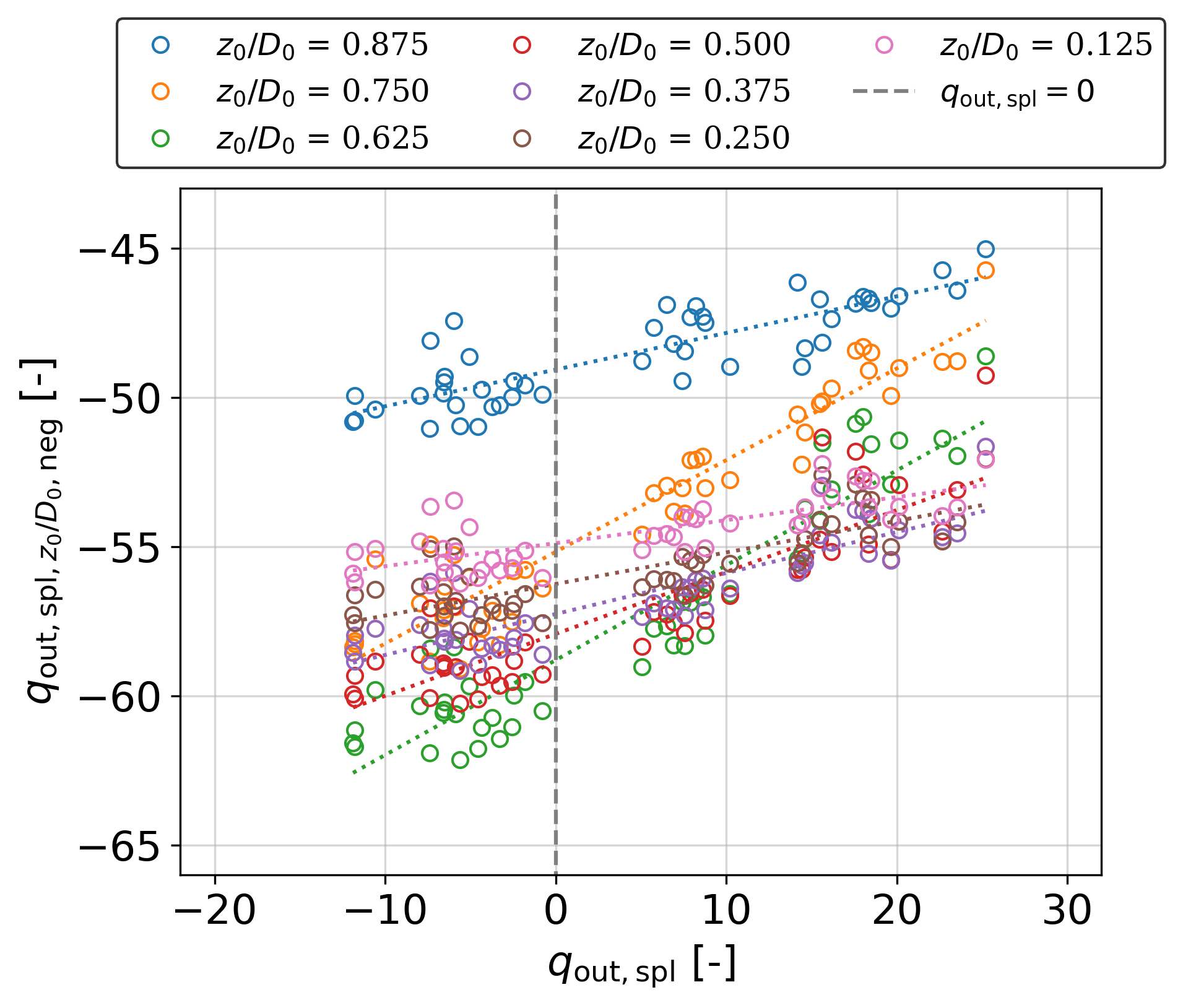}
\label{fig:qout_BP_neg} }
\\
\centering
\subfloat[]{
\includegraphics[width=0.8\textwidth]{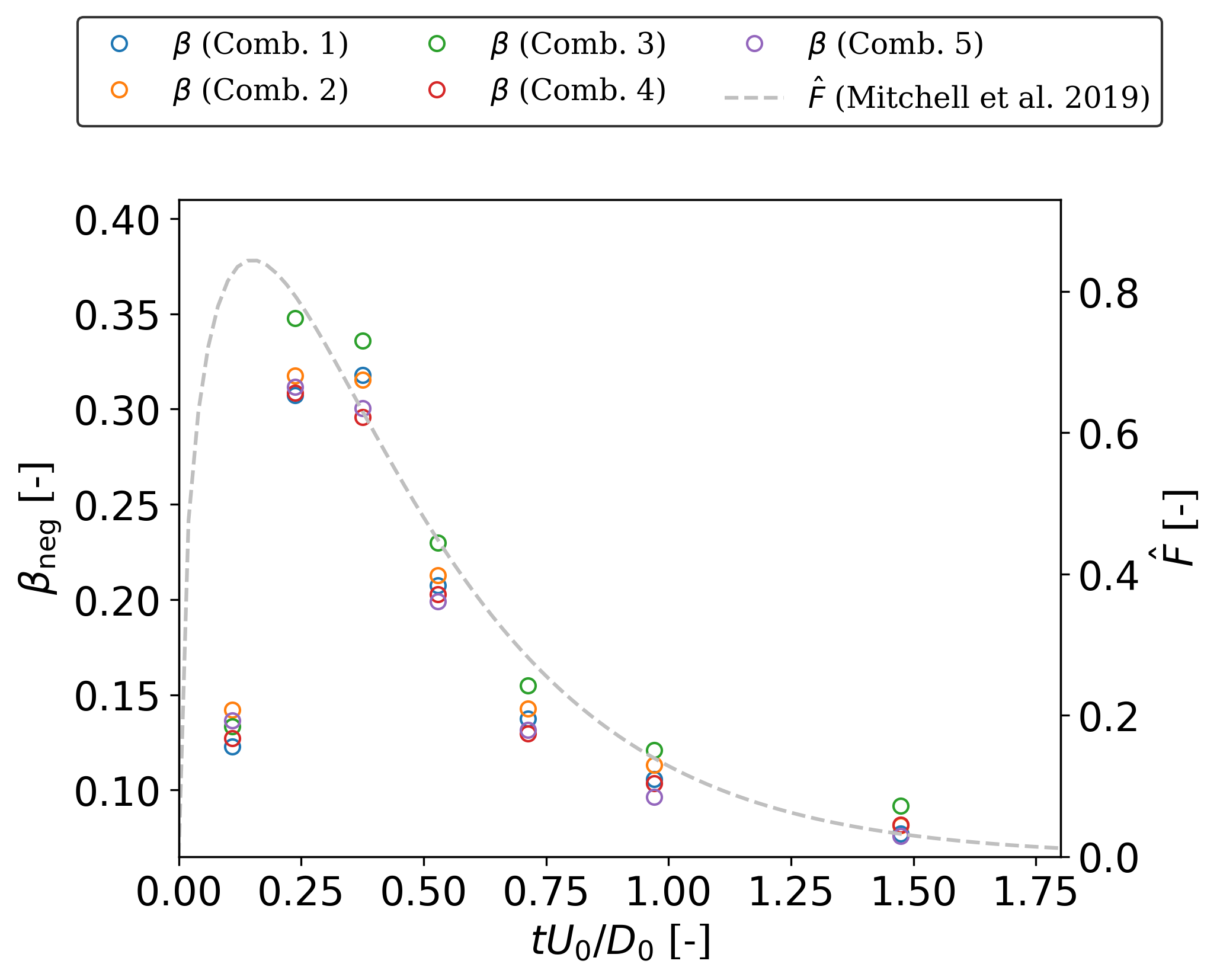}
\label{fig:qGrad_neg}
}
\caption{
\label{fig:qout_BP_qGrad_neg}
(\textit{a}) Contribution of the splashing features to $q_{\mathrm{out,spl},z_0/D_0}$ versus $q_{\mathrm{out,spl}}$ of the test image sequences of combination 1.
(\textit{b}) Slopes of fitted lines $\beta_\mathrm{neg}$ versus normalized impact time $tU_0/D_0$ of all data combinations.
}
\end{figure}

Similar to the trend of $\hat{F}$, after reaching peak values at $tU_0/D_0 = 0.238$ and 0.375, $\beta_{z_0/D_0}$ decreases until $tU_0/D_0 = 0.971$.
However, $\beta$ slightly increases at $tU_0/D_0 = 1.473$, because the lamella of a non-splashing drop develops into a film thicker than that of a splashing drop, as mentioned in \S~\ref{sec:fea_ext}.
In particular, at $tU_0/D_0 = 1.473$, when $z_0/D_0$ decreases to 0.125, this difference in terms of the lamella is the most obvious, and thus could easily be picked up by the FNN.
To check this reasoning, the contributions of splashing and non-splashing features to the computation of $q_{\mathrm{out,spl},z_0/D_0}$ were analysed.
Since the negative values in a $\boldsymbol{w}_{\mathrm{spl},z_0/D_0}$ vector correspond to the features of a splashing drop, the contribution of the splashing features can be computed by summing the products of the elements of $\boldsymbol{w}_{\mathrm{spl},z_0/D_0}$ that have negative values with the corresponding pixel positions in an image.
Similarly, the contribution of the non-splashing features can be computed by summing the products of the elements of $\boldsymbol{w}_{\mathrm{spl},z_0/D_0}$ that have positive values with the corresponding pixel positions in an image.
Thus, let $q_{\mathrm{out,spl},z_0/D_0,\mathrm{neg}}$ and $q_{\mathrm{out,spl},z_0/D_0,\mathrm{pos}}$ be the contributions of the splashing and non-splashing features, respectively, to $q_{\mathrm{out,spl},z_0/D_0}$:
\begin{equation}
q_{\mathrm{out,spl},z_0/D_0} = q_{\mathrm{out,spl},z_0/D_0,\mathrm{neg}} + q_{\mathrm{out,spl},z_0/D_0,\mathrm{pos}}
\label{eq:q_out_pos_neg}.
\end{equation}
The values of $q_{\mathrm{out,spl},z_0/D_0,\mathrm{neg}}$ are plotted against $q_{\mathrm{out,spl}}$ of test image sequences of combination 1 in figure~\ref{fig:qout_BP_neg}.
Only the plots for combination 1 are shown, because similar results were obtained for the other data combinations.
Similar to figure~\ref{fig:qout_BP}, least squares fitting is performed for each $z_0/D_0$ and shown by the dotted lines with the same colours as the respective markers.
Denoted by $\beta_\mathrm{neg}$, the slopes of the fitted lines are plotted against $tU_0/D_0$ for the respective $z_0/D_0$ for all data combinations in figure~\ref{fig:qGrad_neg}.
As can be seen, the value of $\beta_\mathrm{neg}$ fits almost perfectly with the trend of $\hat{F}$, even for $tU_0/D_0 = 1.473$.
This indicates that the splashing features, which include the contour of the main body and the lifted lamella, are dominated by the impact force.
An explanation of the contribution of the non-splashing features can be found in appendix~\ref{app:nonSplashFeatures}.

\section{\label{sec:conclusion} Conclusions and outlook}

The morphological evolution of splashing and non-splashing drops during impact has been compared using an XAI.
An FNN model has been developed as the XAI, comprising a single fully connected layer.
After high classification accuracy had been attained, an analysis of the FNN's classification process was performed.
Feature extraction revealed that the XAI distinguished splashing drops on the basis of a lifted lamella and a higher contour of the main body, while it identified non-splashing drops by a higher asymptotic film thickness of the lamella.
An importance index has been introduced to quantify the contribution of the extracted splashing and non-splashing features to the classification of the XAI model.
The results of this study show that the morphological differences between splashing and non-splashing drops are most pronounced when the impacting drop's apex decreases to 0.750 and 0.625 of the area-equivalent diameter.
The maximum impact force is exerted on the surface immediately before the critical impact time.
The importance index for splashing features closely matches the profile of the normalized impact force.
This shows that the splashing features are most pronounced immediately after the normalized impact force reaches its peak value, rather than at later impact times, when they appear most pronounced to human eyes.
This study has provided an example that clarifies the relationship between the complex morphological evolution of a splashing drop and physical parameters by interpreting the classification of an XAI video classifier.


\backsection[Acknowledgements]{
The authors would like to thank Dr. Masaharu Kameda (Professor, Tokyo University of Agriculture and Technology) and Dr. Yuta Kurashina (Associate Professor, Tokyo University of Agriculture and Technology) for their valuable discussions and suggestions during the lab seminars.
Also, J.Y. would like to express his gratitude to the Epson International Scholarship Foundation and the Rotary Yoneyama Memorial Foundation for their financial support during his PhD study when most of the work was performed.
}

\backsection[Funding]{
This work was funded by the Japan Society for the Promotion of Science (Grant Nos. 20H00222, 20H00223, and 20K20972) and the Japan Science and Technology Agency PRESTO (Grant No. JPMJPR21O5). 
}

\backsection[Declaration of interests]{
The authors report no conflict of interest.}

\backsection[Data availability statement]{
The data that support the findings of this study are available from the corresponding author, Y.T., upon reasonable request.
}

\backsection[Author ORCIDs]{Authors may include the ORCID identifers as follows. 
J. Yee, https://orcid.org/0000-0002-0549-165X
Pradipto, https://orcid.org/0000-0002-9302-767X
A. Yamanaka, https://orcid.org/0000-0002-1781-8330
Y. Tagawa, https://orcid.org/0000-0002-0049-1984
}

\backsection[Author contributions]{
Conceptualization: J.Y., A.Y., and Y.T.;
Data curation: J.Y. and Y.T.;
Formal Analysis: J.Y. and D.I.;
Funding acquisition: Y.T.;
Investigation: J.Y.;
Methodology: all authors;
Project administration: J.Y.;
Resources: Y.T.;
Software: J.Y. and D.I.;
Supervision: A.Y. and Y.T.;
Validation: P., A.Y., and Y.T.;
Visualization: J.Y.;
Writing--original draft: J.Y.; 
Writing--review and editing: all authors.
}

\appendix

\section{Contribution of non-splashing features}\label{app:nonSplashFeatures}

\begin{figure}
\centering
\subfloat[]{
\includegraphics[width=0.45\columnwidth]{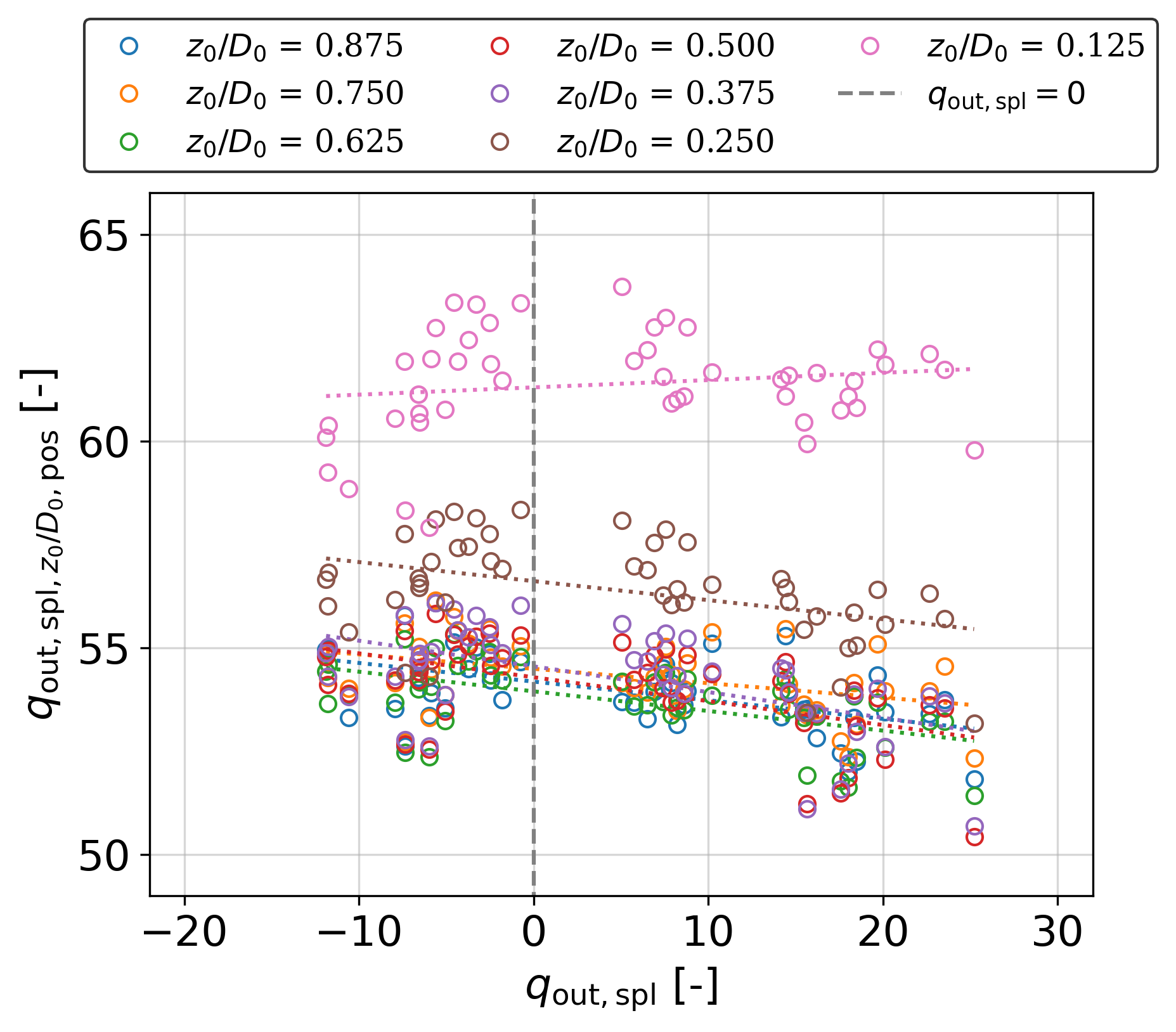}
\label{fig:qout_BP_pos}
}
\subfloat[]{
\includegraphics[width=0.45\columnwidth]{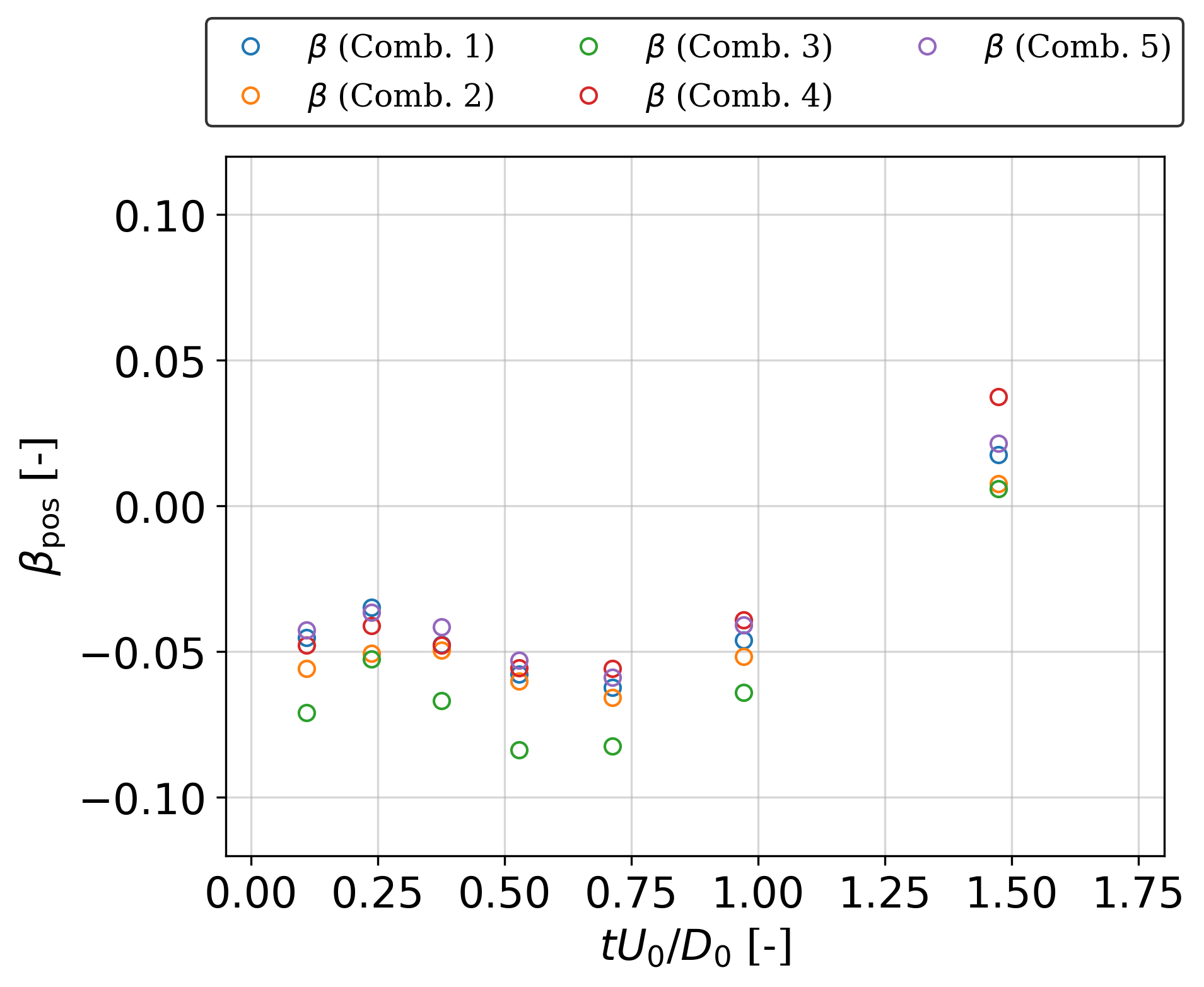}
\label{fig:qGrad_pos}
}
\caption{
\label{fig:qout_BP_qGrad_pos}
(\textit{a}) Contribution of the non-splashing features to $q_{\mathrm{out,spl},z_0/D_0}$ versus $q_{\mathrm{out,spl}}$ from the test image sequences of combination 1.
(\textit{b}) Slopes of fitted lines $\beta_\mathrm{pos}$ versus normalized impact time $tU_0/D_0$ of all data combinations.
}
\end{figure}

The contribution of non-splashing features is explained here.
The values of $q_{\mathrm{out,spl},z_0/D_0,\mathrm{pos}}$ obtained from~\eqref{eq:q_out_pos_neg} are plotted against $q_{\mathrm{out,spl}}$ from test image sequences of combination 1 in figure~\ref{fig:qout_BP_pos}.
Only the plots for combination 1 are shown, because similar results were obtained for the other data combinations.
Similar to figure~\ref{fig:qout_BP}, least squares fitting is performed for each $z_0/D_0$ and shown by the dotted lines with the same colours as the respective markers.
Denoted by $\beta_\mathrm{pos}$, the slopes of the fitted lines are plotted against $tU_0/D_0$ for the respective $z_0/D_0$ for all data combinations in figure~\ref{fig:qGrad_pos}.
As can be seen, the values have much smaller magnitudes than $\beta_\mathrm{neg}$ and are negative except for $tU_0/D_0 = 1.473$.
This indicates that the non-splashing features have less influence on the classification of the FNN than the splashing features.
Besides, the values should not be negative, because this would reduce the classification accuracy.

\begin{figure}
\centering
\subfloat[]{
\includegraphics[width=0.45\textwidth]{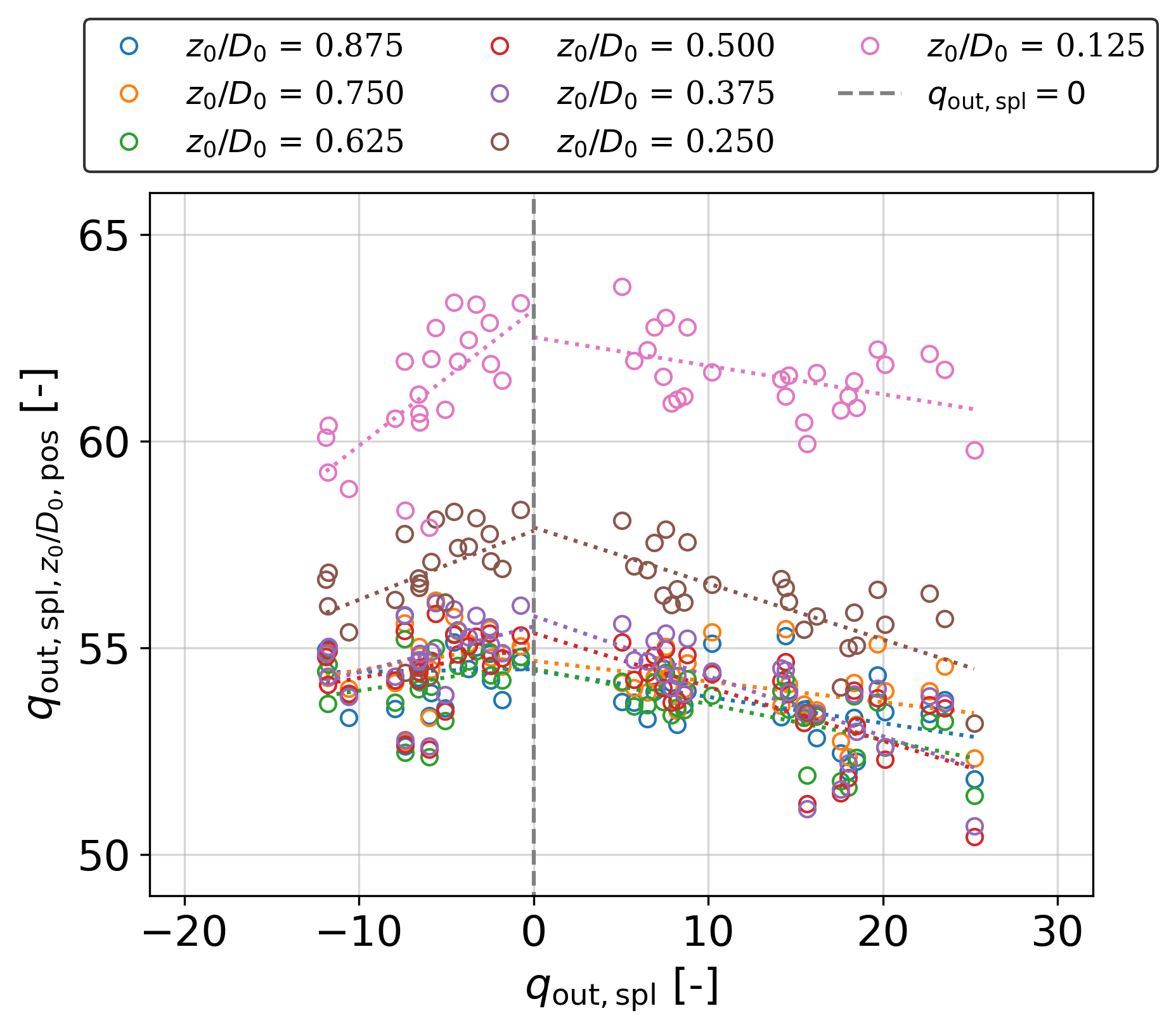}
\label{fig:qout_BP_pos_2}
}
\subfloat[]{
\includegraphics[width=0.45\columnwidth]{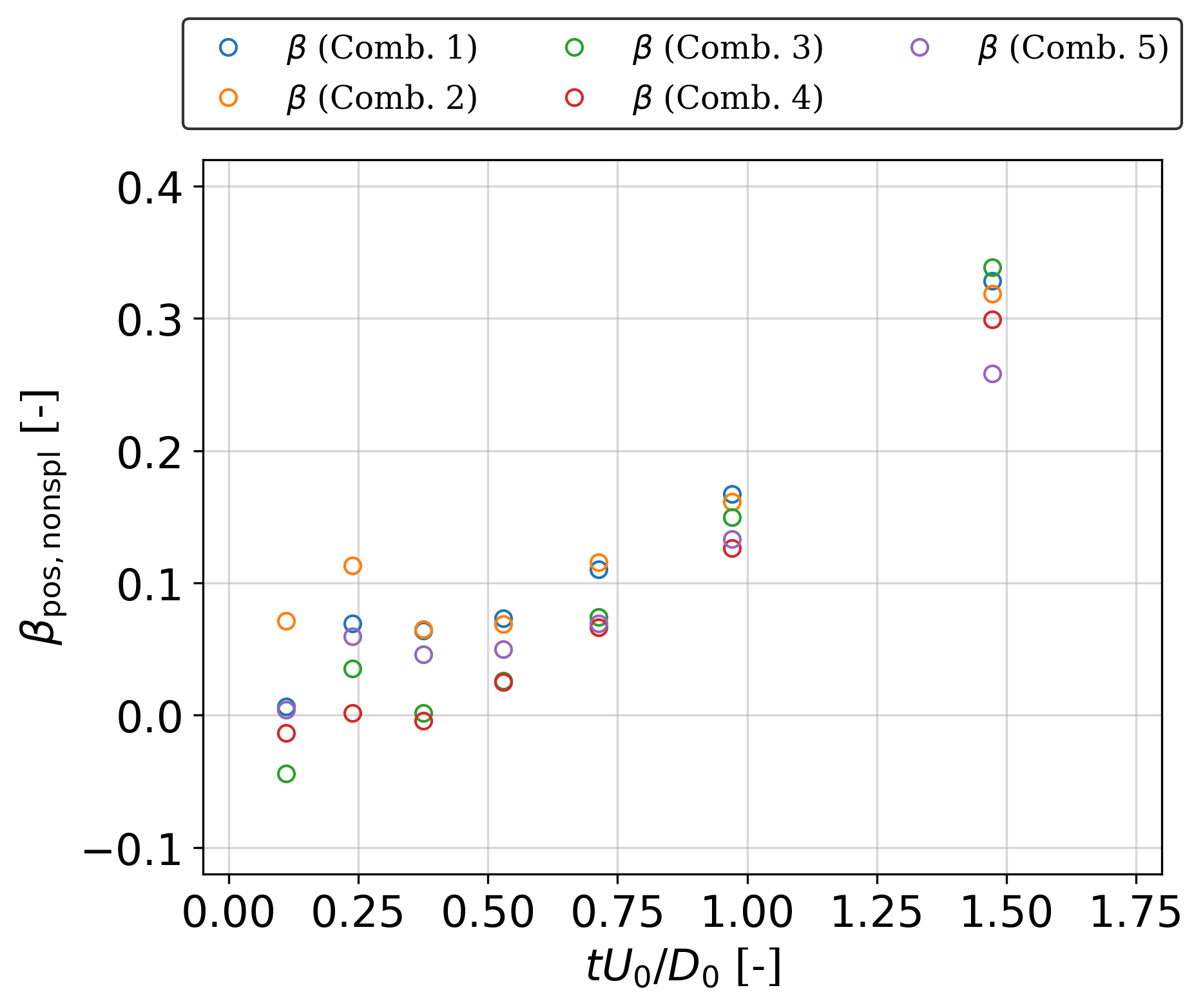}
\label{fig:qGrad_pos_neg_2}
}
\caption{
\label{fig:qout_BP_qGrad_pos_2} 
(\textit{a}) Contribution of the non-splashing features to $q_{\mathrm{out,spl},z_0/D_0}$ versus $q_{\mathrm{out,spl}}$ from test image sequences of combination 1, plotted with two sets of fitted lines: one in the splashing regime where $q_{\mathrm{out,spl},z_0/D_0} > 0$ and the other in the non-splashing regime where $q_{\mathrm{out,spl},z_0/D_0} < 0$.
(\textit{b}) Slopes of the fitted lines in the non-splashing regime.
}
\end{figure}

To explain this, the data markers in figure~\ref{fig:qout_BP_pos} are reproduced in figure~\ref{fig:qout_BP_pos_2}, but this time with two separate least squares fits: one in the splashing regime where $q_{\mathrm{out,spl},z_0/D_0} \geq 0$ and the other in the non-splashing regime where $q_{\mathrm{out,spl},z_0/D_0} < 0$.
As can be seen from figure~\ref{fig:qout_BP_pos_2}, in the non-splashing regime, along with the values of $q_{\mathrm{out,spl}}$, the values of $q_{\mathrm{out,spl},z_0/D_0}$ for all $z_0/D_0$ exhibit an increasing trend, where the slopes of all the fitted lines are positive.
The values of the slopes $\beta_{\mathrm{pos},\mathrm{nonspl}}$ are plotted in figure~\ref{fig:qGrad_pos_neg_2}, from which it can be seen that $\beta_{\mathrm{pos},\mathrm{nonspl}}$ exhibits an increasing trend along with $tU_0/D_0$ until reaching its maximum value at $tU_0/D_0 = 1.473$.
This indicates that although they have less influence on the classification of the FNN than the splashing features, the influence of the non-splashing features increases with $tU_0/D_0$.
As described in \S~\ref{sec:fea_ext}, the main non-splashing feature is the evolution of the lamella of a non-splashing drop from a pre-ejected lamella at the contact line when $tU_0/D_0 = 0.110$ to a film with an asymptotic thickness when $tU_0/D_0 = 0.110$ (see figure~\ref{fig:trained_w}).
The findings here support the argument that the difference between splashing and non-splashing drops in terms of the lamella is greatest at $tU_0/D_0 = 1.473$, and thus could easily be picked up by the FNN.
Since there is no obvious correlation with $\hat{F}$, this indicates that $\hat{F}$ does not have much influence on the evolution of the lamella of a non-splashing drop.
In fact, \citet{lagubeau2012spreading} and \citet{eggers2010drop} reported that the asymptotic film thickness scales with $\Rey^{-2/5}D_0$, and thus the evolution of the lamella of a non-splashing drop is dominated by the viscous force acting on the drop, owing to its low impact velocity $U_0$.
We argue that the value of the viscous force increases with $tU_0/D_0$ as the drop spreads over the surface.
Since there are no data available to support our argument at the moment, we believe that it is important for future work to measure the viscous force that acts on a drop during impact.

Note that the non-splashing and splashing regimes show different trends in figure~\ref{fig:qout_BP_pos_2}.
In the non-splashing regime, when a drop has a higher value of $\Rey$, the asymptotic film thickness is small, as a consequence of which the drop exhibits fewer non-splashing features (i.e., there are fewer elements of $\boldsymbol{w}_{\mathrm{spl},z_0/D_0}$ with positive values that would be cancelled out), which increases the value of $q_{\mathrm{out,spl},z_0/D_0}$.
However, in the splashing regime, along with the values of $q_{\mathrm{out,spl}}$, the values of $q_{\mathrm{out,spl},z_0/D_0}$ for all $z_0/D_0$ exhibit a decreasing trend, with the slopes of all the fitted lines being negative.
This is because, in the splashing regime, the projected area of a splashing drop is larger owing to the ejected secondary droplets and the higher contour of the main body.
Some of this increased area overlaps with the non-splashing features, thus causing the values of $q_{\mathrm{out,spl},z_0/D_0}$ to decrease.


\bibliographystyle{jfm}
\bibliography{jfm}

\end{document}